\documentclass[12pt,preprint]{aastex}

\def\ltsima{$\; \buildrel < \over \sim \;$}
\def\gtsima{$\; \buildrel > \over \sim \;$}
\def\lsim{\lower.5ex\hbox{\ltsima}}
\def\gsim{\lower.5ex\hbox{\gtsima}}
\def\lapp{\ifmmode\stackrel{<}{_{\sim}}\else$\stackrel{<}{_{\sim}}$\fi}
\def\gapp{\ifmmode\stackrel{>}{_{\sim}}\else$\stackrel{<}{_{\sim}}$\fi}
\def\nbss{N_{\rm BSS}}
\def\nhb{N_{\rm HB}}
\def\nrgb{N_{\rm RGB}}

\newdimen\minuswidth    
\setbox0=\hbox{$-$}
\minuswidth=\wd0
\catcode`@=\active
\def@{\kern\minuswidth}
\setbox0=\hbox{\rm0}
 
\shorttitle{Blue Stragglers in NGC2419}
\shortauthors{Dalessandro et al.}
 
\begin{document} 
\title{ Another Non-segregated Blue Straggler Population
  in a Globular Cluster: the
  Case of NGC\,2419\altaffilmark{5}}

\author{
E. Dalessandro\altaffilmark{1,2},
B. Lanzoni\altaffilmark{1},
F.R. Ferraro\altaffilmark{1},
F. Vespe\altaffilmark{2}, 
M.Bellazzini\altaffilmark{3},
R.T. Rood\altaffilmark{4}}

\altaffiltext{1}{Dipartimento di Astronomia, Universit\`a degli Studi
di Bologna, via Ranzani 1, I--40127 Bologna, Italy}
\altaffiltext{2}{ASI,Centro di Geodesia Spaziale, contrada Terlecchia,
I-75100, Matera, Italy}
\altaffiltext{3}{INAF--Osservatorio Astronomico di Bologna, via
Ranzani 1, I--40127 Bologna, Italy}
\altaffiltext{4}{Astronomy Department, University of Virginia,
P.O. Box 400325, Charlottesville, VA, 22904}
\altaffiltext{5}{Based on observations with the NASA/ESA {\it HST}, obtained at the
Space Telescope Science Institute, which is operated by AURA, Inc., under
NASA contract NAS5-26555. Also based on  SUBARU observations collected
  at the Hawaii National Astronomical
Observatoty of Japan.}

\begin{abstract}
We have used a combination of {\it ACS-HST} high-resolution and
wide-field {\it SUBARU} data in order to study the Blue Straggler Star
(BSS) population over the entire extension of the remote Galactic
globular cluster NGC\,2419.  The BSS population presented here is
among the largest ever observed in any stellar system, with more than
230 BSS in the brightest portion of the sequence. The radial
distribution of the selected BSS is essentially the same as that of
the other cluster stars. In this sense the BSS radial distribution is
similar to that of $\omega$ Centauri and unlike that of all Galactic
globular clusters studied to date, which have highly centrally
segregated distributions and, in most cases, a pronounced upturn in the
external regions. As in the case of $\omega$ Centauri, this evidence
indicates that NGC\,2419 is not yet relaxed even in the central
regions. This observational fact is in agreement with estimated
half-mass relaxation time, which is of the order of the
cluster age.
\end{abstract} 


\section{INTRODUCTION}
This paper is part of a series  aimed at studying the complex
interplay between dynamics and stellar evolution in a sample of
globular clusters (GCs) with different structural properties. To this
purpose we are using the so-called Blue Stragglers Star (BSS)
population as a probe. BSS are the most abundant and
common population of stars which significantly deviates from the
evolutionary path defined by a simple, old stellar population. In the
Color Magnitude Diagram (CMD), BSS define a sparsely populated
sequence more luminous and bluer than the Turn-Off point (TO) of a
normal hydrogen-burning main-sequence (MS). Hence they appear as stars
more massive (see also Shara et al. 1997) and younger than the bulk of
the cluster population. They are thought to form from the evolution
of binary systems, either via mass-transfer/coalescence phenomena in
primordial binaries \citep[PB-BSS;][]{mtbss1, mtbss2}, or by stellar
mergers induced by collisions \citep[COL-BSS;][]{colbss}.  Since
collisional stellar systems like GCs dynamically evolve over a
time-scale significantly shorter than their age, BSS are expected to
have sunk into the cluster core, as have the majority of the most massive
objects (like binaries and other binary by-products) harbored in the
system.

In many GCs the projected radial distribution of BSSs has been found
to be bimodal: highly peaked in the center, with a clear-cut dip at
intermediate radii, and with an upturn in the external regions. Such a
behavior has been confirmed in at least 7 GCs: M3, 47 Tuc, NGC\,6752,
M5, and M55, NGC\,6388 (all references in Dalessandro et al. 2007) and
M53 (Beccari et al. 2008). Dynamical simulations (Mapelli et al. 2006;
Lanzoni et al. 2007a,c) suggest that the observed central peak is
mainly due to COL-BSS formed in the core and/or PB-BSS sunk into the
center because of dynamical friction, while the external rising branch
is made of PB-BSS evolving in isolation in the cluster outskirts.
Even in those GCs that do not show any bimodality the BSS always
appear to be significantly more segregated in the central regions than
the reference cluster stars.  The only exception to these general
observational features is $\omega$ Centauri (hereafter $\omega$ Cen):
the large population of BSS discovered by Ferraro et al. (2006,
hereafter F06) in this giant stellar system has the same radial
distribution of the normal cluster stars. This is a clear evidence
that $\omega$ Cen is not fully relaxed, even in the central regions,
and therefore, the dynamical evolution of the cluster has not
significantly altered the radial distribution of these stars. It is
likely that the vast majority of BSS observed in this cluster are the
progeny of primordial binaries evolved in isolation (see also Mapelli
et al. 2006).

Here we direct our attention to another massive cluster (NGC2419) which
shares a number of properties with $\omega$ Cen. This 
remote object ($d \sim 81$\,kpc; Harris et al. 1997)
is one of the most luminous clusters in the Galaxy ($M_V = -9.4$;
see Bellazzini 2007, herefater B07), similar to $\omega$ Cen and M54
(NGC\,6715). It has been suggested that both of the latter clusters 
are the remnants of stripped cores of dwarf spheroidals (see, e.g.,
Layden \& Sarajedini 2000; Bekki \& Freeman 2003).
With its high luminosity and half-light radius ($r_h \simeq 25$\,pc;
B07), NGC\,2419 lies (together with $\omega$ Cen and M54) in the
$(r_h,~M_V)$ plane well above the locus defined by all the other
Galactic GCs. Indeed, it is the most significant outlier, thus
suggesting that it also might be the stripped core of a former dwarf
galaxy (van den Bergh \& Mackey 2004; Mackey \& van den Bergh 2005).
Further, Newberg et al. (2003) suggested that NGC\,2419
could be somehow connected with the Sagittarius (Sgr) dwarf
spheroidal, since it seems to be located in a region with an
overdensity of type-A stars which is in the same plane as the
tidal tails of Sgr.
However, the high-quality CMDs of NGC\,2419 recently published by
Ripepi et al. (2007, hereafter R07; see also B07) do not show any
evidence of multiple stellar populations, in contrast to $\omega$ Cen
(Lee et al. 1999; Pancino et al. 2000; Bedin et al. 2004; Rey et
al. 2004; Sollima et al.  2005) and possibly M54 (Layden \& Sarajedini
2000; see also Monaco et al. 2005). It is however possible that for
such a metal-poor cluster ($[Fe/H]=-1.97$; Ferraro et al. 1999b), the
range in metallicities for the sub-population components is so small
that different sequences cannot be seen in the CMD (Mackey \& van den
Bergh 2005; Federici et al. 2007).

In order to further investigate the dynamical status and the stellar
populations of this remote cluster, here we present a multi-wavelength
study of BSS in NGC\,2419. By combining HST high-resolution data,
with wide-field SUBARU images, we sampled the total radial extension of the
cluster.  This allowed us to study and compare the projected
radial distributions of BSS and other cluster stars in different
evolutionary stages.  The data and photometric reductions are
described in \S~\ref{sec:data}.  A general overview of the CMD is discussed in \S~\ref{sec:cmd}. The
BSS population is described in \S~\ref{sec:BSS}, and the Discussion is
presented in \S~\ref{sec:discus}.

\section{OBSERVATIONS AND DATA ANALYSIS}
\label{sec:data}
\subsection{The data sets}
To study the
crowded cores of high-density systems and simultaneously cover the
total cluster extensions,  a combination of high
resolution observations of the central regions and complementary
wide-field images is needed.

\begin{figure}[!hp]
\begin{center}
\includegraphics[scale=0.7]{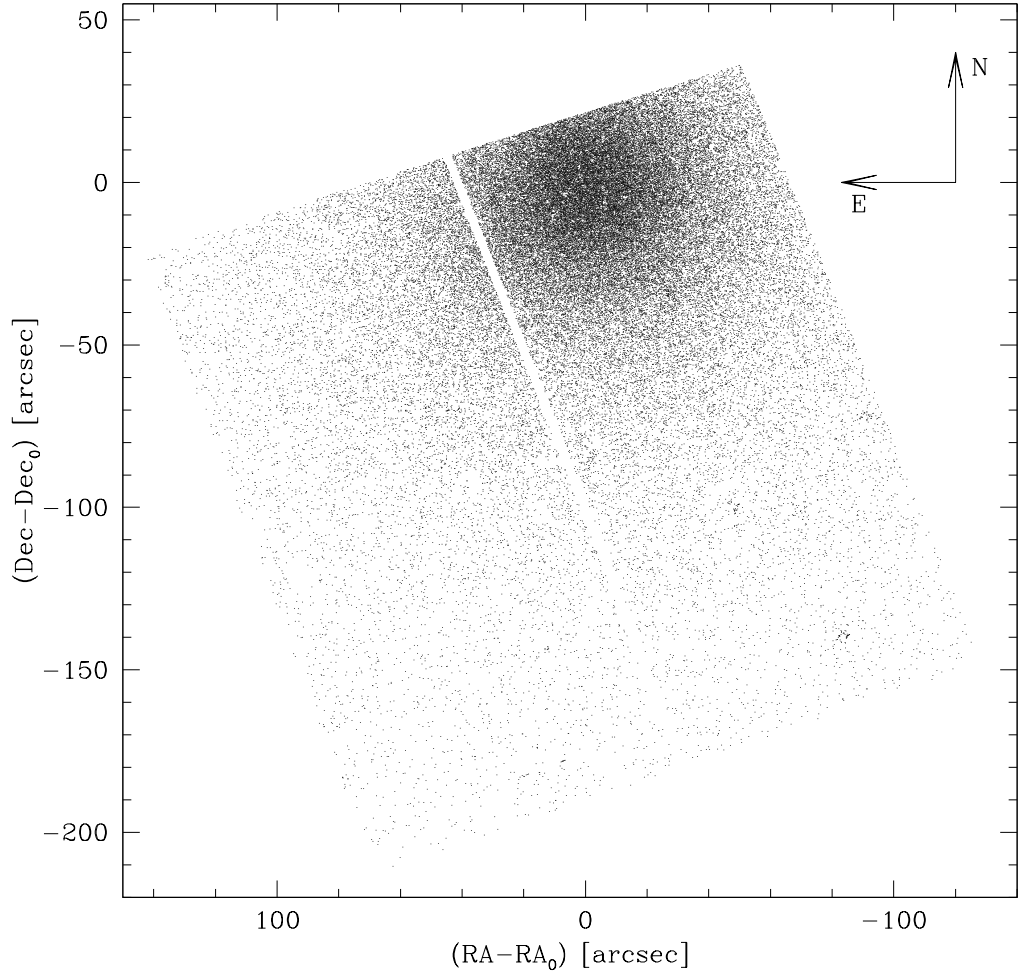}
\caption{Map of the {\it HST sample}. $RA_0$ and $Dec_0$ are
the right ascension and the declination of the center of gravity
of the cluster.}
\label{acs}
\end{center}
\end{figure}

\emph{1. High resolution set --} This is composed of a series of
public images obtained with the Wide Field Channel of the Advanced
Camera for Surveys (ACS) on board the Hubble Space Telecope (HST): two
F435W ($\sim B$ filter) images with $t_{\rm exp}=800$\,sec each, two
F555W ($\sim V$ filter) images with $t_{\rm exp}=720$\,sec, and two
F814W ($\sim I$ filter) images with $t_{\rm exp}= 676$\,sec (Prop
GO9666, P.I. Gilliland). These are the highest resolution ($\sim
0.05\arcsec\,{\rm pixel}^{-1}$) observations available to date for
NGC\,2419. Unfortunately the ACS images are off-centered (see
Figure\,\ref{acs}), and they do not completely sample the most central
region of the cluster.  As in previous works (see, e.g.,
Dalessandro et al. 2007), average ACS images were obtained in each
filter, and they were corrected for geometric distorsion and effective
flux (Sirianni et al. 2005).  The data reduction has been performed
using the ROMAFOT package (Buonanno et al. 1983), specifically
developed to perform accurate photometry in crowded regions (Buonanno
\& Iannicola 1989).

\begin{center}
\begin{figure}[!p]
\includegraphics[scale=0.7]{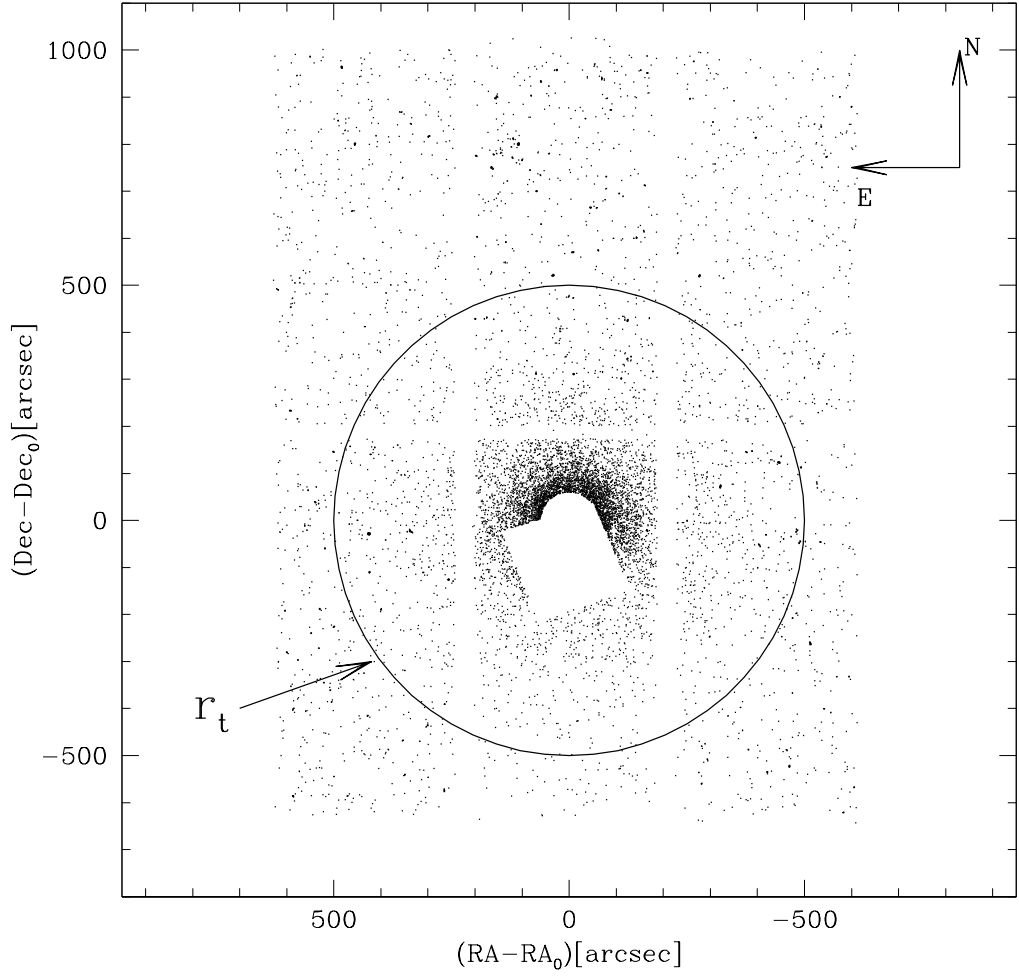}
\caption{Map of the {\it SUBARU sample}. The circle with radius
  $r_t=500\arcsec$ (adopted as tidal radius) centered in the cluster
  center is shown as a solid line.}
\label{subaru}
\end{figure}
\end{center}

\emph{2. Wide field set --} We have used a set of public $V$ and $I$
images obtained with the SUBARU Prime Focus Camera (Suprime-Cam) of
the 8.2\,m SUBARU telescope at the Hawaii National Astronomical
Observatoty of Japan. The Suprime-Cam is a mosaic of ten
$2048\times4096$ CCDs, which covers a $34'\times27'$ field of view
(FoV) with a pixel scale of $0.2\arcsec$. A combination of
long-exposures ($t_{\rm exp}=180$\,sec) and median exposure ($t_{\rm
  exp}=30$\,sec) images has been retrieved from the Subaru Archive Web
site (SMOKA).  As shown in Figure\,\ref{subaru}, the cluster is
centered in the chip \#2 and it is totally included in the five
adjacent chips; therefore only these six chips have been considered in
the present study.  We have applied standard pre-reduction procedures
(correction for bias, flat-field and overscan) using
IRAF{\footnote{IRAF is distributed by the National Optical Astronomy
    Observatory, which is operated by the Association of Universities
    for Research in Astronomy, Inc., under cooperative agreement with
    the national Science Foundation}} tools. The reduction was
performed independently for each image using the PSF fitting software
DoPhot (Schechter et al. 1993).
         
\subsection{Astrometry, center of gravity and photometric calibration}

The ACS and SUBARU data have been placed on the absolute astrometric
system by using the stars in common between each single chip and the
SDSS data set used by B07, that, in turn, was astrometrized on the
GSC-II astrometric reference star catalog.  Hundreds of stars have
been matched in each chip, thus allowing a very precise determination
of the stellar absolute positions.  The resulting
rms residuals (a measure of the internal
astrometric accuracy) were of the order of $\sim 0\farcs 3$ both in
Right Ascension ($\alpha$) and Declination ($\delta$).
  
The photometric calibration of the ACS catalog has been performed in
the VEGAMAG stystem using the relations and zero-points described in
Sirianni et al. (2005).  Then, the SUBARU catalog has been homogenized
to the ACS one.  In order to transform of the instrumental SUBARU
magnitudes into the ACS VEGAMAG system, a subsample of a few hundred
stars in common between the SUBARU and the ACS FoVs has been selected,
and the following relations have been obtained:
\begin{equation}
I_{\rm ACS}-i_{\rm SUBARU} = -0.55 \times (V-I)_{\rm ACS} +27.41
\end{equation}
\begin{equation}
V_{\rm ACS}-v_{\rm SUBARU} = -0.20 \times (V-I)_{\rm ACS} +27.46
\end{equation}
where $i_{\rm SUBARU}$ and $v_{\rm SUBARU}$ are the instrumental $I$
and $V$ magnitudes in the SUBARU sample referred to 1 second exposure.
In this way a final list of absolute positions and homogeneous
(VEGAMAG) magnitudes for all the stars in the two catalogs was
obtained.

In order to determine the Center of Gravity ($C_{\rm grav}$) of the
cluster, we have computed the barycenter of all the stars found in the
ACS catalog at a distance $r<10\arcsec$ from the center quoted by
Harris (1996). A circular region of $10\arcsec$ radius is the
maximum available area completely covered by the ACS observations (see
Fig.\,\ref{acs}). The absolute positions ($\alpha$, $\delta$) of the
stars have been averaged using an iterative technique 
described in previous works (e.g., Montegriffo et al. 1995; Ferraro et
al. 2003).  We have excluded stars brighter than $V=19.5$ since they
are saturated in the ACS images. The same procedure has been
repeated for three different magnitude cuts ($V<24$, $V<23.5$, and
$V<23$) in order to check for any possible statistical or
spurious fluctuations. The three measures agree within $\sim 1\arcsec$
and their mean value has been adopted as best estimate of $C_{\rm
  grav}$: $\alpha = 7^{\rm h}\,38^{\rm m}\,8\fs 47^s$ and $\delta=38^{\circ}\,52'\,
55\farcs 0$,
with an uncentainty of $0\farcs 5$ in both $\alpha$ and $\delta$.
This new determination is in agreement with that listed by
Harris (1996).

Given the coordinates of $C_{\rm grav}$, we have divided the dataset
in two main samples: the \emph{HST sample}, which includes all the
stars found in the ACS catalog, and the \emph{SUBARU sample}, that
consists of stars not included in the ACS FoV and lying at
$r>60\arcsec$ from the cluster center.  The latter choice implies that
a small region (a segment of a circle located $\sim 20\arcsec$ North
from the cluster center) is covered neither by the HST nor by the
SUBARU sample (see Fig.\ref{subaru}). This  conservative choice is made
to avoid incompletness effects of the ground based
observations in the most crowded central regions of the
cluster.\footnote{However, note that the annular region between
  $20\arcsec$ and $60\arcsec$ from the cluster center is well sampled
  (at $\sim 70\%$) by the ACS sample.}

\section{CMD overall characteristics and the HB morphology}
\label{sec:cmd}
\begin{figure}[!p]
\begin{center}
\includegraphics[scale=0.7]{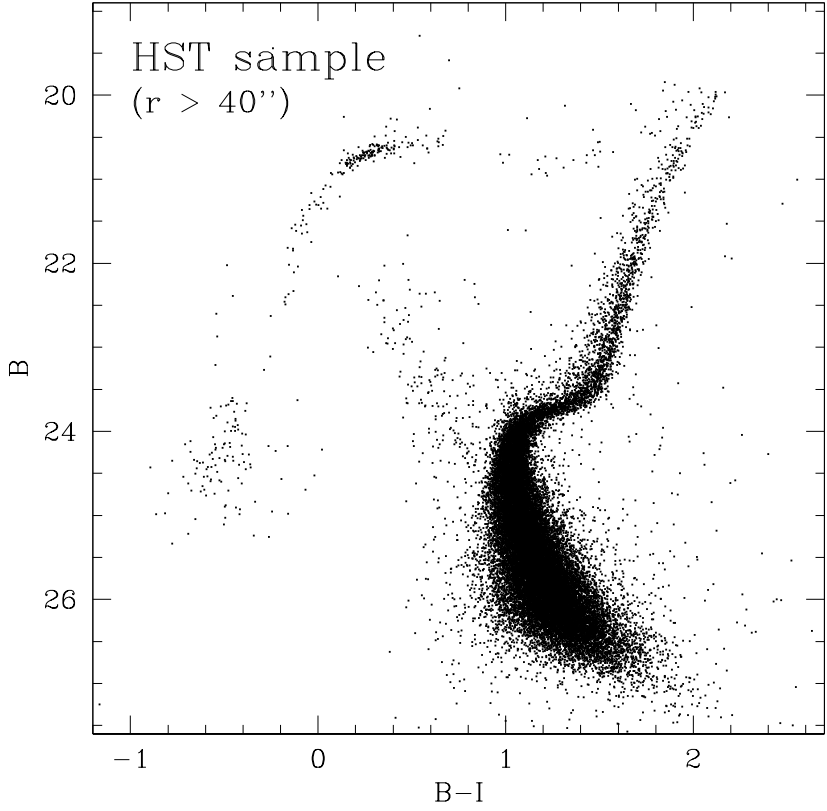}
\caption{$(B,\,B-I)$ CMD of the {\it HST sample} for $r>40\arcsec$
  from the center, reaching down $B \sim 27$.}
\label{bbi}
\end{center}
\end{figure}

The CMD of stars in the HST sample is shown in Figure\,\ref{bbi}.
This is the deepest CMD ever published for NGC\,2419, reaching down to
$B\sim 27$. All the main cluster evolutionary sequences are clearly
defined and well populated.  The stars in the brightest ($B<19.5$)
portion of the red giant branch (RGB) are not shown in the figure,
because they are heavily saturated in these exposures.  The MS-TO of
the cluster is located at $B_{\rm TO}=24.5\pm 0.1$.  Particularly
notable is the horizontal branch (HB) morphology, which looks quite
complex, with a long HB blue tail (BT) extending well below the
cluster MS-TO. The peak of the HB population is located at $B\sim20.7$
and $(B-I)\sim 0.2$. The HB population significantly decreases with
decreasing luminosity along the BT. A poorly populated region (a gap?)
is visible at $B\sim 23.4$, separating the extreme extention of the BT
and a clump of stars extending down to $B\sim 25$.  Following the
nomenclature adopted in Dalessandro et al. (2007), these are extreme
HB (EHB) stars with the faintest ones being Blue Hook (BHk)
stars. Definitive assignment to these groups will require UV
photometry.

\begin{figure}[!p]
\begin{center}
\includegraphics[scale=0.7]{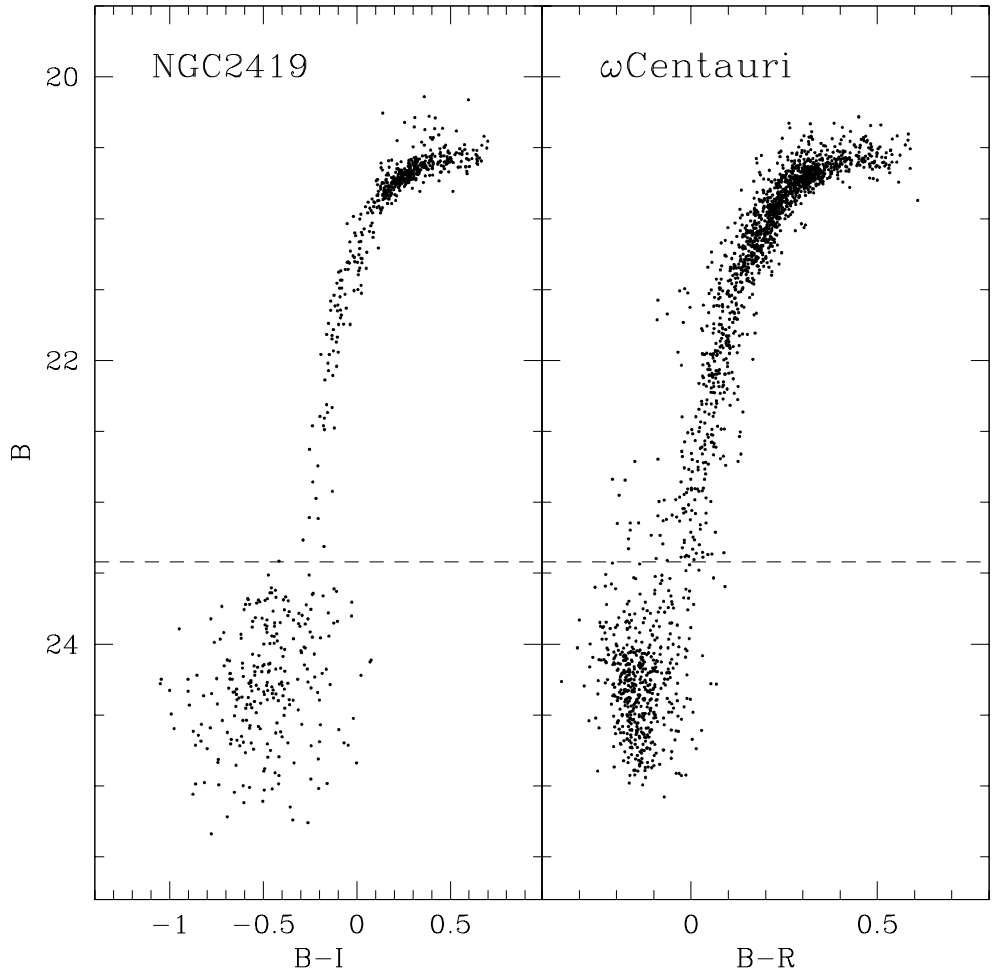}
\caption{Comparison of the HB morphology of NGC\,2419 and $\omega$ Cen
  (from F06). Only stars in the ACS FoV are plotted. The HB of
  $\omega$ Cen has been shifted by $\delta B = 5.6$ to match that of
  NGC\,2419. The dashed line marks the brightest boundary of the BHk
  population. }
\label{hb}
\end{center}
\end{figure}

In Figure\,\ref{hb} we show a direct comparison between the HB of
NGC\,2419, and that of $\omega$ Cen (from Ferraro et al. 2004),
suitably shifted (by $\sim 5.6$ magnitudes) in order to match the HB
level of NGC\,2419. The two HBs show very similar extension and
morphology. The only significant difference is that EHB/BHk stars in
NGC\,2419 are much more spread-out in color $\delta(B-I)\sim 1$ than
the same population in $\omega$ Cen. The rms scatter between the
magnitude measurements in the two single images,  
is $\sigma_B\sim 0.1$ and $\sigma_I\sim 0.24$\,mag in the $B$ and
$I$ bands, respectively.  Thus the
photometric error in $(B-I)$ is $\sigma_{B-I}\sim 0.26$\,mag at the
level of BHk stars. The observed color spraed is about 4$\sigma$ and
may thus be real. 
To demonstrate more clearly
the striking similiarity of these HBs, in Figure \,\ref{hbdistr} we
show the normalized  $B$ magnitude distribution of HB stars.  
The percentage of stars in
three portions of the branch is also designated in the figure. Beyond
general appearance, the HBs are quantitatively similar: {\it (i)} both
the HBs extend for almost 4.5 mag; {\it (ii)} both the distributions
show a well defined peak, an extended tail and a EHB/BHk clump; {\it
(iii)} the bulk of the HB population ($\sim 58\%$) is localized in the
brightest 1 magnitude portion of the branch; {\it (iv)} the BT
is 10--12\% of the population; {\it (v)}
both the EHB/BHk clumps extend for roughly 1.5 magnitudes and they
comprise $\sim 30\%$ of the total HB population.

As discussed in Dalessandro et al. (2007), the nature of BHk stars is
still unclear: they may be related to the so-called late hot flashers
\citep[][Catelan 2007]{moehler2808}, or due to high helium abundances
(as suggested by Busso et al. 2007, in the case of NGC6388; see also
Caloi \& D'Antona 2007; D'Antona et al. 2005), or related to the
evolution of binary systems \citep{heber02}. However, the detection of
a population of BHk stars in a low-metallicity cluster as NGC\,2419
clearly demonstrates that the process producing these extremely hot HB
stars can efficiently work in any metallicity environment: NGC\,6388
($[Fe/H]\sim -0.4$), NGC\,2808 ($[Fe/H]\sim -1.1$), $\omega$ Cen
($[Fe/H]\sim -1.6)$, M54 ($[Fe/H]\sim -1.8)$ and NGC\,2419
($[Fe/H]\sim -2$). It is interesting to note
that NGC\,2419 is very massive as are the other BHk
clusters. We have also checked the EHB/BHk radial distributions with
respect to the brightest portion of the HB and the RGB.  The
significance of the difference has been quantified with a
Kolmogorov-Smirnov (KS) test: the radial distribution of the BHk
population is consistent with that of normal cluster stars, in
agreement with similar findings in NGC\,6388 (Rich et al. 1997,
Dalessandro et al. 2007) and $\omega$ Cen (Ferraro et al. 2004). However,
the evidence presented in Sect.\,\ref{sec:radBSS} demonstrate that
NGC\,2419 is not relaxed even in the central regions, hence no
segregation is expected for these stars even in the case they were
binaries. Moreover, as discussed in Dalessandro et al. (2007), it is
important to remember that the lack of segregation of the EHB/BHk
population is not a firm proof of the non-binarity of EHB/BHk stars,
since they could be low-mass binaries, with a total mass similar (or
even lower) than ``normal'' cluster stars (for example, a
0.5\,$M_\odot$ He-burning star with a 0.2\,$M_\odot$ He white dwarf
companion).

\subsection{Density profile and distance modulus estimate}
\label{sec:kingdist}
\begin{figure}[!p]
\begin{center}
\includegraphics[scale=0.7]{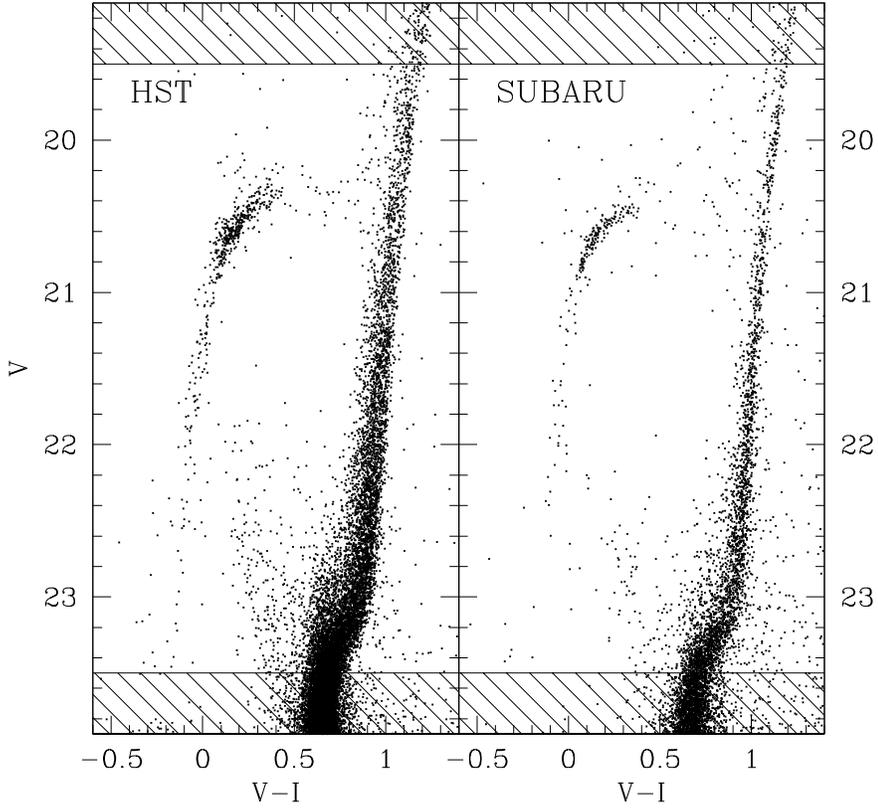}
\caption{CMDs used to derive the surface density profile of
  NGC\,2419. The hatched regions indicate stars that have been
  excluded because they are saturated in the {\it HST sample} (those
  with $V<19.5$), or in order to avoid incompleteness effects (stars
  fainter than $V=23.5$).}
\label{vvi}
\end{center}
\end{figure}

The $(V,~V-I$) CMDs of the HST and SUBARU samples defined in Section
2.2 are shown in Figure\,\ref{vvi}.  Thanks to the high-resolution ACS
images of the cluster core and the wide FoV of the SUBARU
observations, we have properly sampled the stellar population over the
entire cluster extension.  We have then used this data-set to
determine the projected density profile of NGC\,2419 using direct star
counts, from $C_{\rm grav}$ out to about $1000\arcsec$.

\begin{figure}[!p]
\begin{center}
\includegraphics[scale=0.7]{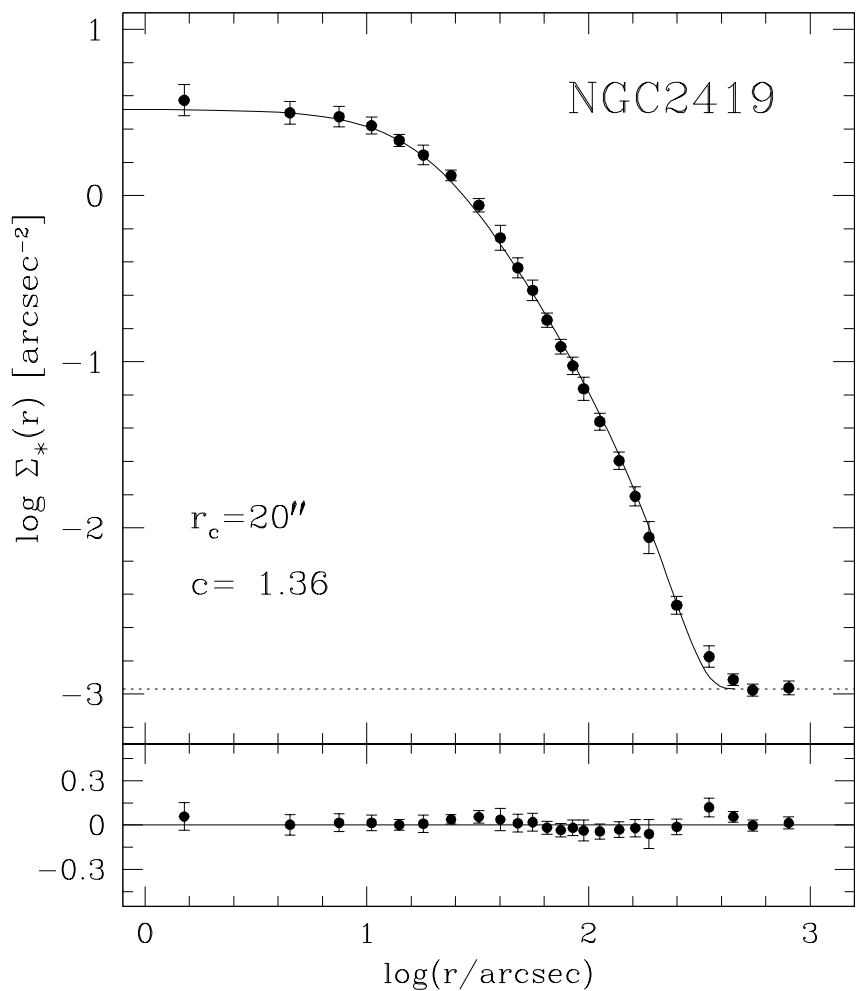}
\caption{Observed surface density profile ({\it dots and error bars})
  and best-fit King model ({\it solid line}). The radial profile is in
  units of number of stars per square arcseconds.  The {\it dotted
    line} indicates the adopted level of the background (corresponding
  to $\sim 4$\,stars/arcmin$^2$), and the model characteristic
  parameters (core radius $r_c$ and concentration $c$) are marked in
  the figure. The lower panel shows the residuals between the
  observations and the fitted profile at each radial coordinate.}
\label{king}
\end{center}
\end{figure}

Stars with $V< 19.5$ are saturated in the ACS sample and therefore
have been excluded from the analysis; however, since they are small in
number, this produces a negligible effect on the global result.  In
order to avoid incompleteness biases we have also excluded stars
fainter than $V=23.5$. Using the same procedure described in Ferraro
et al. (1999a) the whole sample has been divided in 24 concentric
annuli, each centered on $C_{\rm grav}$ and suitably split in a number
of subsectors.  The number counts have been calculated in each
subsector and the corresponding densities were obtained dividing them
by the sampled area (taking into account the incomplete spatial
coverage of the region between $20\arcsec$ and $60\arcsec$).  The
stellar density of each annulus has then be defined as the average of
the subsector densities and its standard deviation is computed from
the variance among the subsectors. The resulting projected surface
density profile is plotted in Figure\,\ref{king}. As apparent, the
outermost two points show a flattening of the stellar number density,
and their average (corresponding to $\sim 4$\,stars/arcmin$^2$) has
therefore been used as an estimate of the background contribution.
The derived radial density profile is well fit by an isotropic
single-mass King model (solid line in Fig.\,\ref{king}), with
concentration $c=1.36$ and core radius $r_c=20\arcsec$, yielding a
``formal'' value of the cluster tidal radius of $r_t \sim 460\arcsec$
and a half-mass radius of $r_h\sim 58\arcsec$. These parameters are
essentially equal to those obtained by B07 and in good agreement with
other previous determinations (see, e.g., Table 2 in B07).

\begin{figure}[!p]
\begin{center}
\includegraphics[scale=0.7]{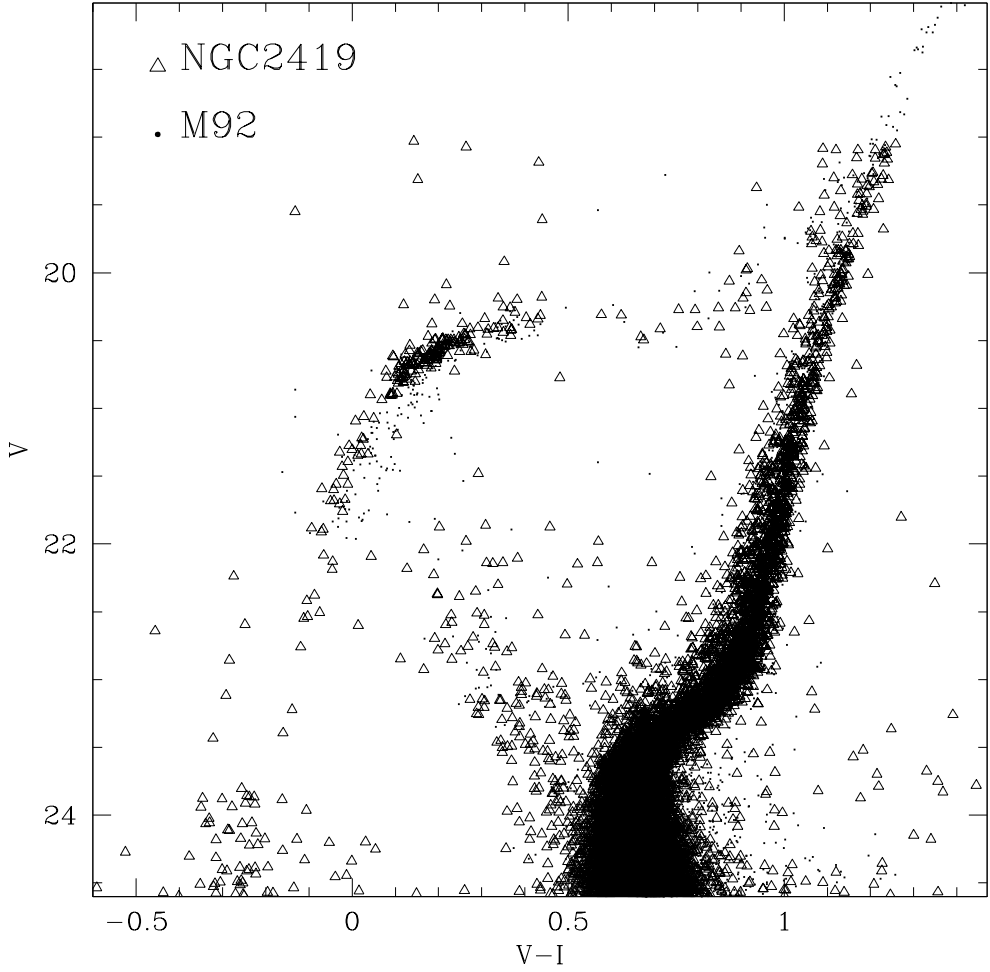}
\caption{$(V,\,V-I)$ CMD of M92 ({\it dots}) superimposed onto that of
  NGC\,2419 ({\it triangles}) after a magnitude shift of $\delta
  V=5.25$ and a color shift of $\delta(V-I)=0.14$.}
\label{m92}
\end{center}
\end{figure}

We have used the available high-quality data set also for deriving an
independent estimate of the distance to NGC\,2419. To do this, we
compared the CMD shown in Fig.\,\ref{bbi}, to that of M92 (NGC\,6341),
one of the ``prototype'' Galactic GCs, with similar metallicity
([Fe/H]$\sim -2$, Ferraro et al. 1999b).  We have used a combination
of WFPC2 and ACS data of M92 (F.R. Ferraro et al. 2008, in
preparation), obtained through filters F555W ($\sim V$) and F814W
($\sim I$).  We have shifted the CMD of M92 onto that of NGC\,2419
until a good match between the main evolutionary sequences (RGB, HB,
sub-giant branch and TO region) of the two clusters was reached (see
Figure\,\ref{m92}).  This has required a color shift $\delta
(V-I)=0.14$ and $\delta V =5.25$, similar to that obtained by Harris
et al. (1997) from an analogous comparison based on independent data
sets. Figure\,\ref{m92} shows that a really nice matching of all the
evolutionary sequences of the two clusters can be achieved. This
evidence also suggests that the two clusters have a similar age (in
agreement with Harris et al. 1997, who estimated an age difference of
$\sim 1$\,Gyr for the two objects).

By assuming the distance modulus $(m-M)_0 = 14.78$ and the reddening
$E(B-V)=0.02$ for M92 (Ferraro et al. 1999b), and by using the standard
absorption coefficient ($A_V=3.1$ and $A_I=1.7$), we have obtained
$E(B-V)=0.12\pm0.03$ and $(m-M)_V = 20.09$, corresponding to a true
distance modulus $(m-M)_0=19.72$, for NGC\,2419. The reddening
obtained from this procedure is in good agreement with the value
derived by Harris et al. (1997), who quoted $E(B-V)=0.11$, and it is
also agreement with the value $E(B-V)=0.08$ adopted by R07 within the
errors. Taking a conservative estimate of $\sigma\sim 0.1$\,mag, we
finally adopt $(m-M)_0=19.7\pm 0.1$.
This yields a real distance $d\simeq 87 \pm 4$\,kpc. Within the
uncertanties, this estimate is in agreement with both that found by
Harris et al. (1997; $d=81\pm 2$\,kpc) and that obtained by R07 using
the mean luminosity of the RR Lyrae stars ($d=83.2 \pm
1.9$\,kpc). Assuming this distance, the physical dimension of the core
radius and of the half-mass radius of the cluster can be obtained:
given $r_c=20\arcsec$ and $r_h=58\arcsec$ (see above), we obtain
$r_c=8.4$\,pc and $r_h=24.5$\,pc, respectively.  By adopting the total
integrated magnitude $V_t=10.47$ quoted by B07, the absolute cluster
magnitude is $M_V=-9.6$. This value, combined with the
size of the the half-mass radius, confirms the anomalous
position of NGC\,2419 in the $r_h$ versus $M_V$ plane (van den Bergh
\& Mackey 2004).

\section{THE POPULATION OF BSS}
\label{sec:BSS}

\subsection{Population selection}
\label{sec:popBSS}
\begin{figure}[!p]
\begin{center}
\includegraphics[scale=0.7]{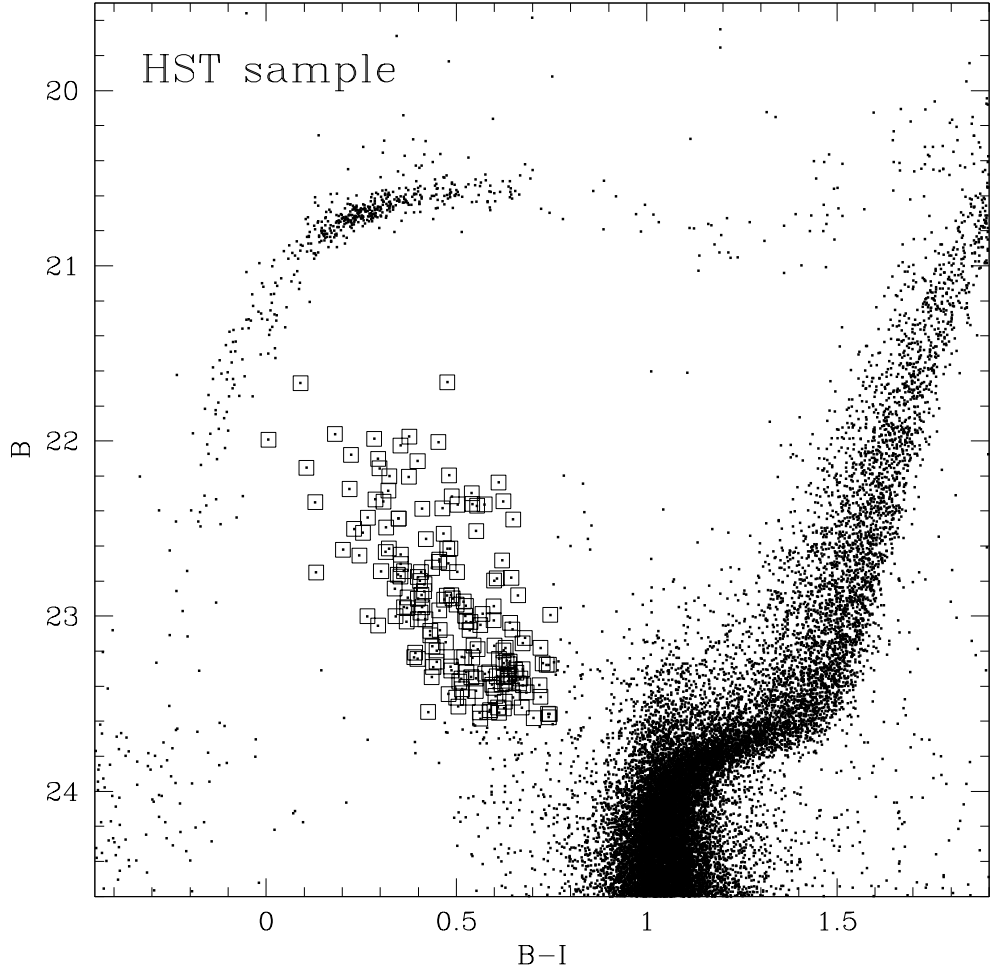}
\caption{BSS population (squares) selected in the $(B,\,B-I)$ CMD of
  the {\it HST sample}.}
\label{bbisel}
\end{center}
\end{figure}

\begin{figure}[!p]
\begin{center}
\includegraphics[scale=0.7]{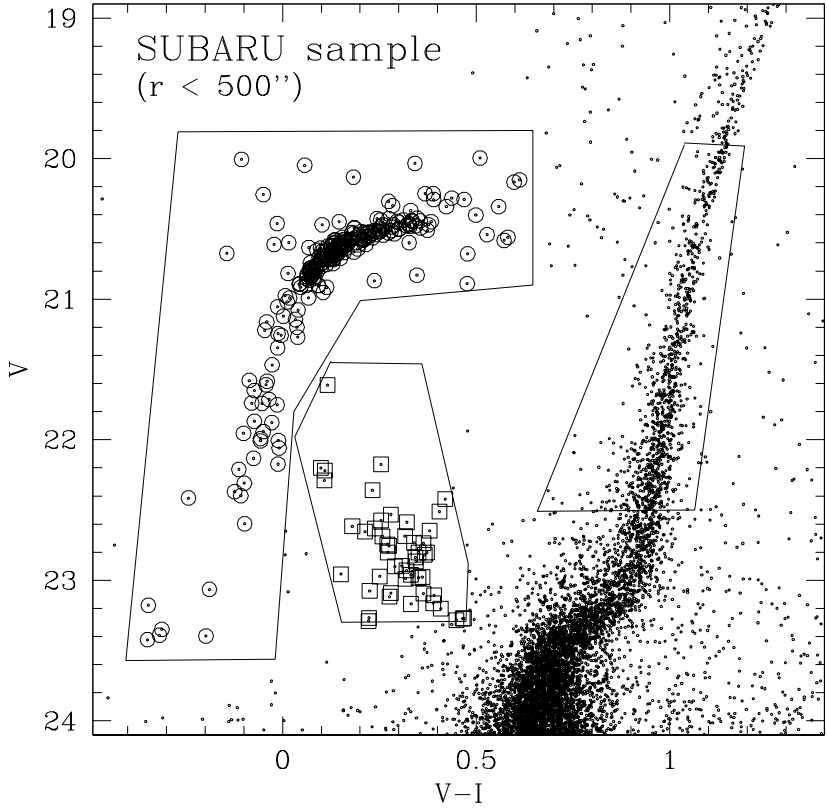}
\caption{$(V,\,V-I)$ CMD of the SUBARU sample for $r<500\arcsec$ from
  the center, with the adopted BSS, HB and RGB selection boxes
  highlighted. The selected BSS and HB populations are also marked
  with open squares and circles, respectively.}
\label{vvisel}
\end{center}
\end{figure}

To select the BSS population we have chosen to use the ($B,~B-I$) CMD,
in which the BSS sequence is better defined. To avoid spurious effects
due to sub-giant branch star blends and Galaxy field star
contamination, only stars brighter than $B\simeq 23.6$ (corresponding
to $\sim$ 1 mag above the TO) and with $B-I<0.75$ have been selected
(see Figure\,\ref{bbisel}). The resulting number of candidate BSS in
the \emph{HST sample} is 183. The position of the bulk of these stars
in the ACS ($V,~V-I$) CMD has then been used to define the BSS
selection box for the \emph{SUBARU sample}. This is shown in
Figure\,\ref{vvisel}, with the faint and red edges corresponding to
$V\simeq 23.3$ and $V-I<0.48$, respectively.  The resulting number of
candidate BSS found in the entire SUBARU sample is 67, out of which 49
are found within the ``safe'' distance of $\sim 500\arcsec$ from the
cluster center. This distance is sligthly larger than the ``formal''
tidal radius obtained in Sect.\,\ref{sec:kingdist} and takes into
account possible uncertanties in the determination of the latter.  The
positions and magnitudes of the all the 232 candidate BSS thus
selected are listed in Table 1, which is available in full size in
electronic form.\footnote{Several SX Phoenicis variables have been
  found by R07. However a direct comparison between these stars and
  our BSS sample is not possible, since the R07 catalog is not yet
  published.}

Reference populations representative of the ``normal'' cluster stars
are needed to properly study the BSS radial distribution. We 
considered both the HB and the RGB. Since the HST and the SUBARU
samples have the $V$ and $I$ filters in common, we performed a
homogeneous selection of these populations in the $(V,~V-I)$ plane.
The HB selection box (see Fig.\,\ref{vvisel}) has been drawn 
to limit the contribution of field contamination in the
bright-red portion of the sequence (i.e., we have required that $V-I < 0.65$
at $V\sim 20$) and in order to exclude the EHB/BHk clump ($V \gsim 23.6$).
The EHB/BHk stars populate a region located $\sim 1$ magnitude below
the MS-TO (see Fig.\,\ref{m92}), which is very close to the detection
limit of the $V$ and $I$ observations. Thus, they could be severely
affected by incompletness bias, and we have therefore preferred not to
include them in the HB reference population. However, since their
radial distribution is indistinguishable from that of the other HB
stars, this exclusion has negligible effect on the following
results. The total number of HB stars thus selected within
$500\arcsec$ is 765, with 528 found in the HST sample and 237 in the
SUBARU one.
The RGB population has been selected along the RGB mean ridge line
between $V\simeq 19.9$ and $V\simeq 22.5$ (see the selection box in
Fig.\,\ref{vvisel}).  This choice has been dictated by the fact that
the brightest portion of the RGB sequence is saturated in the HST
sample, and its faintest portion is contaminated by Galactic field
stars, especially in the SUBARU sample. The total number of these
stars within $500\arcsec$ is 3250, with 2337 found in the HST sample
and 913 in the SUBARU one.

\subsection{BSS radial distribution}
\label{sec:radBSS}
\begin{figure}[!p]
\begin{center}
\includegraphics[scale=0.7]{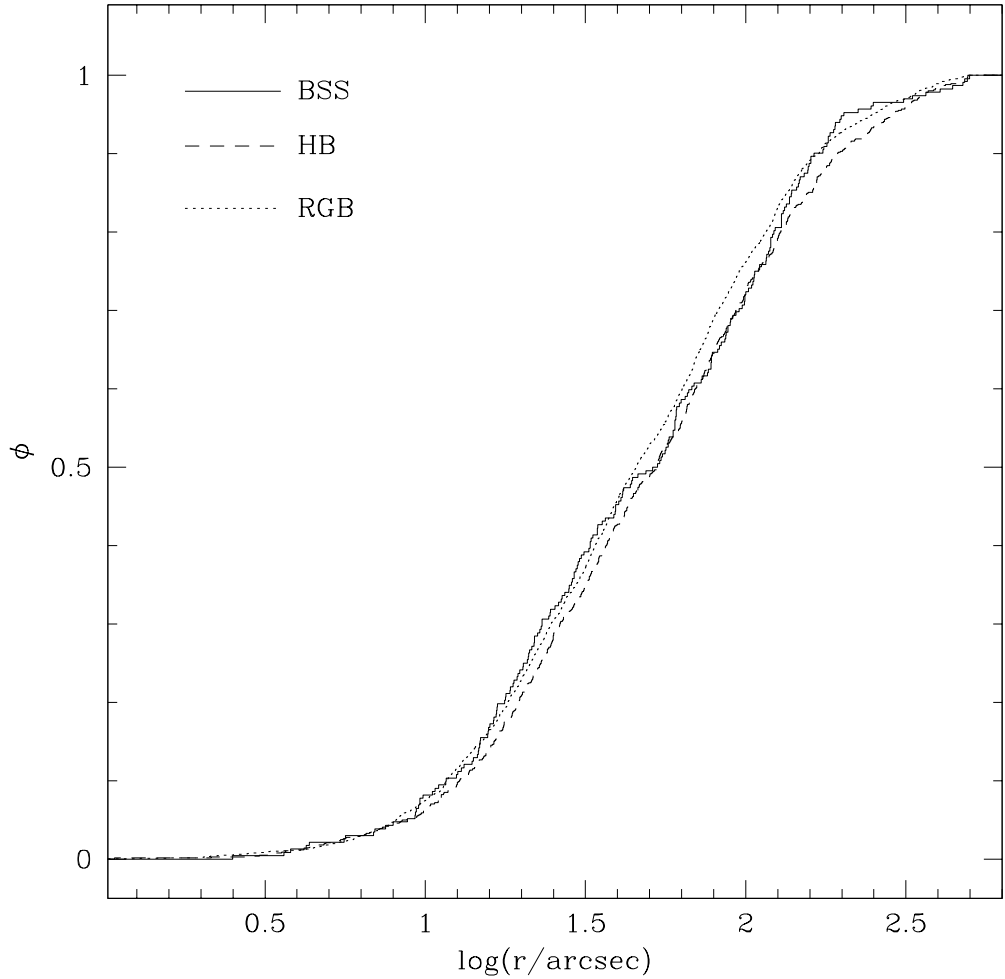}
\caption{Cumulative radial distribution of BSS (solid line), HB
  (dashed line) and RGB stars (dotted line) as a function of the
  projected distance from the cluster center, for the combined
  HST+SUBARU sample at $r<500\arcsec$. The populations show
  essentially the same radial distribution.}
\label{KS}
\end{center}
\end{figure}

A first qualitative comparison between the cumulative radial
distribution of BSS and that of the reference populations (see
Figure\,\ref{KS}) has been performed using the KS test. This gives 
70\% and 50\% probabilities that the BSS population is extracted
from the same population as the HB and RGB stars, respectively.  Hence
there is preliminary evidence that the radial distribution of BSS is
indistinguishable from that of the ``normal'' cluster population, in
contrast to what found in most of the typical GCs (see references in
Dalessandro et al. 2007).

For a more detailed analysis, we have used the same technique
described in previous works (see, e.g., F06). The sampled area within
$r=500\arcsec$ has been divided in 5 concentric annuli centered on
$C_{\rm grav}$.  In each of these we have counted the number of BSS,
HB and RGB stars.  However, the examination of the external regions
($r>500\arcsec$) of the SUBARU sample CMD suggests that the selected
(BSS, HB, and RGB) populations can be affected by contamination from
stars in the Galactic field. In order to account for this effect we
adopted the statistical correction as used in previous papers (see,
e.g., Dalessandro et al. 2007). To do this we selected a rectangular
region of $\sim 70\, {\rm arcmin}^2$ located at $r>650\arcsec$, i.e.
well beyond the formal tidal radius of the cluster.  The CMD of this
region clearly shows that the Galaxy field population is dominant
relative to the cluster one. Then we counted the number of stars in
this region lying in the BSS, HB and RGB selection boxes showed in
Fig.\,\ref{vvi} and derived the following values of the field star
densities: $D^{\rm field}_{\rm BSS}= 0.03\,{\rm stars\,arcmin^{-2}}$,
$D^{\rm field}_{\rm HB}= 0.06\,{\rm stars\,arcmin^{-2}}$ and $D^{\rm
  field}_{\rm RGB}= 0.14\,{\rm stars\,arcmin^{-2}}$. These quantities
allow us to estimate the impact of the field contamination on the
selected samples: 6 BSS ($\sim 2\%$), 12 HB ($\sim 1.5\%$) and 31
RGB($\sim 1\%$), essentially all in the most external annulus, could
be field stars (see Table\,\ref{tab:counts}).  Though the effect of
the field contamination is small, in the following we use the
statistically decontaminated samples in order to determine the
population ratios and the radial distribution.

By using the King model, the distance modulus and the reddening
estimated in Sect.\,\ref{sec:kingdist}, the luminosity sampled in each
annulus ($L^{\rm samp}$) has also been estimated.  Then for each
annulus we have computed the double normalized ratio defined in
Ferraro et al. (1993):
\begin{equation}
R_{\rm pop}=\frac{N_{\rm pop}/ N^{\rm tot}_{\rm pop}}{L^{\rm samp}/
  L^{\rm samp}_{\rm tot}
},
\label{eq:Rpop}
\end{equation}
with pop= BSS, HB and RGB.  We find that $R_{\rm HB}$ and $R_{\rm
  RGB}$ are essentially constant and close to unity (see $R_{\rm HB}$
in Figure\,\ref{Rpop}).  This is what expected for any post-MS
population according to the stellar evolution theory (Renzini \& Fusi
Pecci 1988). Surprisingly we also find that the double normalized
ratio of BSS is constant, and it is fully consistent with the reference
populations (Fig.\,\ref{Rpop}).  Using the number counts listed in
Table\,\ref{tab:counts}, we have also computed the specific
frequencies $\nbss/\nhb$, $\nbss/\nrgb$ and $\nhb/\nrgb$.  We find
that all these ratios are almost costant all over the entire extension
of the cluster (see Figure\,\ref{Npop}), confirming again that no
signatures of mass segregation are visible for the BSS population of
NGC\,2419.\footnote{Note that this result is
  independent of which portion of the RGB is selected. In fact, 
  similar results are
  obtained also by considering the RGB in the same magnitude range of
  the BSS.}

\section{DISCUSSION}
\label{sec:discus}
In most previously surveyed Galactic GCs (M3; 47 Tuc, NGC\,6752,M5,
M55, NGC\,6388, M53) the BSS radial distribution has been found to be
bimodal (higly peaked in the core, decreasing to a minimun at
intermediate radii, and rising again in the external regions).  The
mechanisms leading to bimodal radial distributions have been studied
for some clusters using dynamical simulations (see Mapelli et
al. 2004, 2006; Lanzoni et al. 2007a,c): the observed central peak is
mainly made up of collisionally formed BSS and/or PB-BSS sunk into the
core because of dynamical friction; the external rising branch is
composed of PB-BSS evolving in isolation in the cluster outskirts.  In
contrast, the BSS radial distribution of NGC\,2419 is essentially the
same as that of the other ``normal'' stars in the cluster. Previously,
$\omega$ Cen was the only GC known to have a flat BSS radial
distribution.  F06 (see also Meylan \& Heggie 1997) argued that
two-body relaxation had not led to the complete relaxation of $\omega$
Cen even in the central core. Our result here suggests that the same
situation holds for NGC\,2419.

We can compare this observational evidence with theoretical
time-scales expected on the basis of the cluster structural
parameters.  Following equation (10) of Djorgovski (1993), we computed
the cluster central relaxation time ($t_{r_c}$) by adopting
$m=0.3\,M_{\odot}$ for the average stellar mass, and $M/L=3$ for the
mass-to-light ratio and $M_{V\odot}=4.79$ for the $V$ band solar
magnitude. The integrated magnitude obtained in
Sect.\,\ref{sec:kingdist} then leads to a total cluster mass of $1.7
\times 10^6 M_{\odot}$, and a total number of stars of $5.7\times
10^6$. By assuming $\rho_0 \simeq 25 \,M_\odot\,{\rm pc}^{-3}$ (Pryor
\& Meylan 1993), and given the value of the core radius
($r_c=8.4$\,pc) derived in Sect.\,\ref{sec:kingdist}, we obtain
$t_{r_c}\sim 6$\,Gyr, which is about half the cluster age ($t=12$--13
Gyr; Harris et al. 1997). Thus some evidence of mass segregation
should be visible at least in the core, at odds with the observed flat
distribution of BSS. We can also compute the characteristic relaxation
time-scale for stars as massive as BSS ($M_{\rm BSS}\sim 1.2
M_{\odot}$; see F06) at the cluster half-mass radius ($r_h$) using
equation (10) of Davies et al. (2004). Since $r_h\sim 24.5$\,pc (see
Sect.\,\ref{sec:kingdist}) we obtain $t_{r_h}\sim 18$\,Gyr, thus
suggesting that no significant segregation is expected for stars as
massive as the BSS in the outer parts of the clusters, in agreement
with our observational results.  This result is similar to that found
for $\omega$ Cen by F06 where the relaxation time in the core was
found to be $\sim$ half the cluster age.  In the case of $\omega$ Cen
a number of possible explanations were examined; for instance, the
possibility that $\omega$ Cen is the relic of a partially disrupted
galaxy, which was much more massive in the past. A similar argument
can be advocated for NGC\,2419, which has also been suspected to be
the relic of a small dwarf galaxy interacting with the Milky Way (van
den Bergh \& Mackey 2004; Federici et al. 2007 and references
therein).  However, it should be noted that the above estimates of the
relaxation times are rough, and since the predicted value of $t_{r_c}$
is only a factor of 2 smaller than the cluster age, more detailed
computations are needed before further interpret these results.

\begin{figure}[!p]
\begin{center}
\includegraphics[scale=0.7]{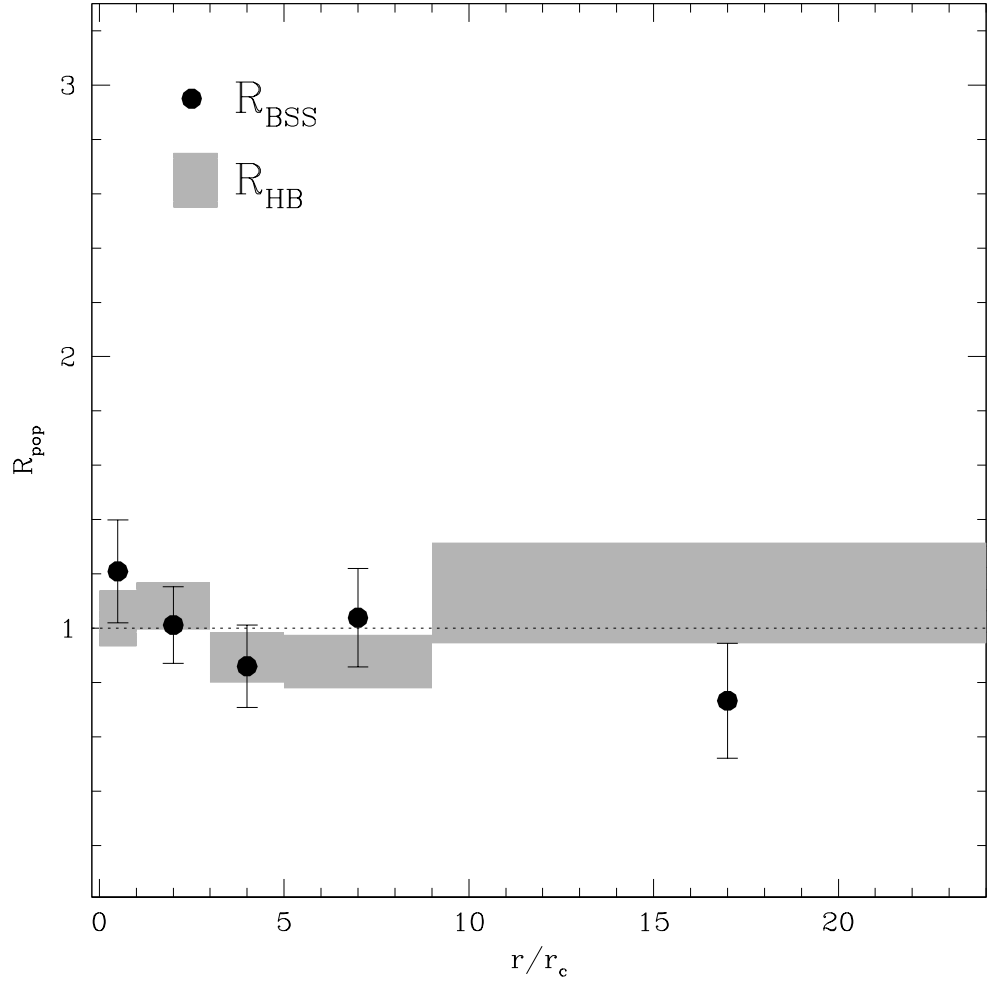}
\caption{Double normalized ratios, as defined in eq. (\ref{eq:Rpop}),
  of BSS (dots) and HB stars (grey rectangular regions), plotted as a
  function of the radial coordinate expressed in units of the core
  radius.  The vertical size of the grey rectangles corresponds to the error
  bars.  The two distributions are almost constant around unity over
  the entire cluster extension, as expected for any normal,
  non-segregated cluster population.  }
\label{Rpop}
\end{center}
\end{figure}

\begin{figure}[!p]
\begin{center}
\includegraphics[scale=0.7]{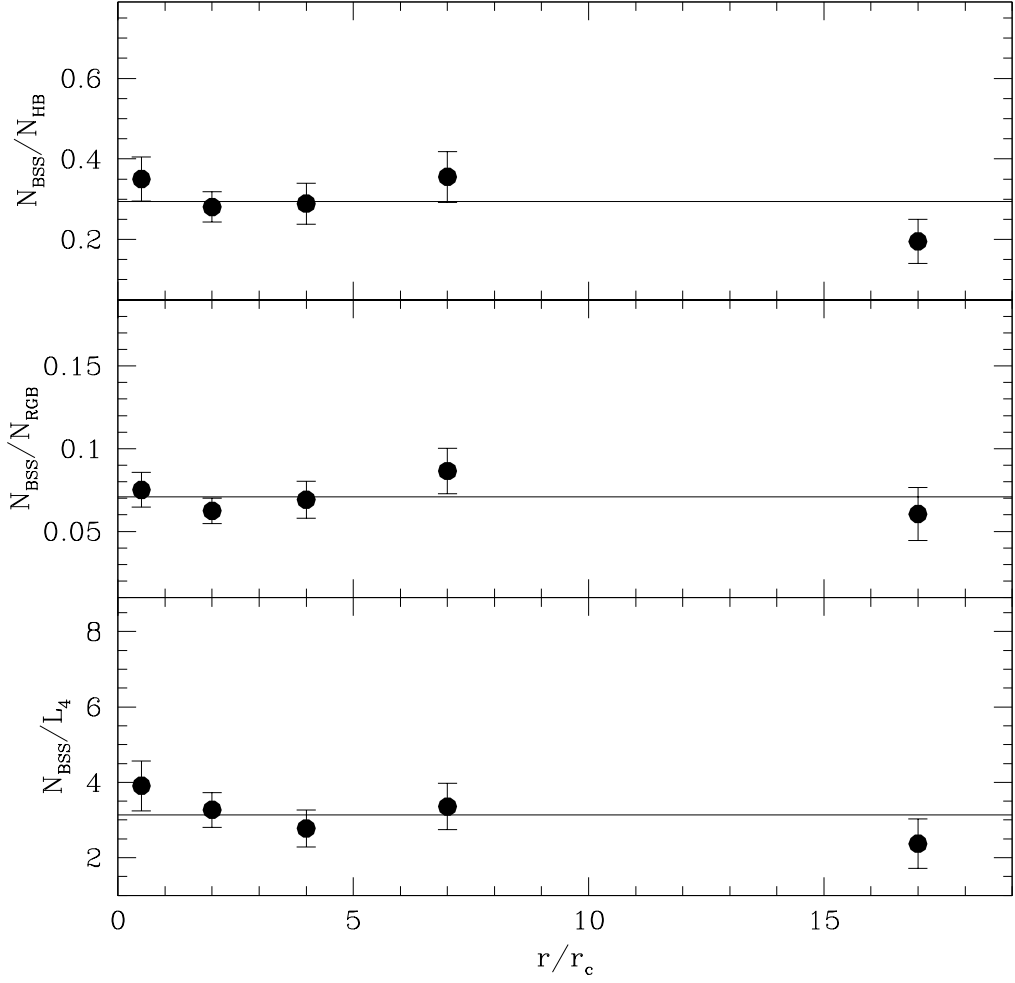}
\caption{Specific frequency of BSS with respect to HB stars (upper
  panel), RGB stars (middle panel), and the sampled luminosity in
  units of $10^4 L_{\odot}$ (bottom panel) as a function of the
  projected distance from the cluster center in units of $r_c$.}
\label{Npop}
\end{center}
\end{figure}

From the observational side, the BSS radial distribution shown in
Figs.\,\ref{Rpop} and \ref{Npop} suggests that in NGC\,2419 (as in
$\omega$ Cen) stellar collisions have played a minor (if any) role in
modifying the radial distribution of massive objects and probably also
in generating exotic binary systems. If dynamical evolution plays a
central role in NGC\,2419, the observed flat BSS distribution can be
explained only by invoking an {\it ad hoc} formation/destruction rate
balancing the BSS population in the core and in the outer region of
the cluster. It is more likely that this flat distribution arises
because the BSS we are observing result from the evolution of
primordial binaries whose radial distribution has not been altered by
the dynamical evolution of the cluster.  Thus, as in the case of
$\omega$ Cen, the BSS population observed in NGC\,2419 could be a pure
population of PB-BSS, and it can be used to evaluate the incidence of
such a population in stellar systems.

\begin{figure}[!p]
\begin{center}
\includegraphics[scale=0.7]{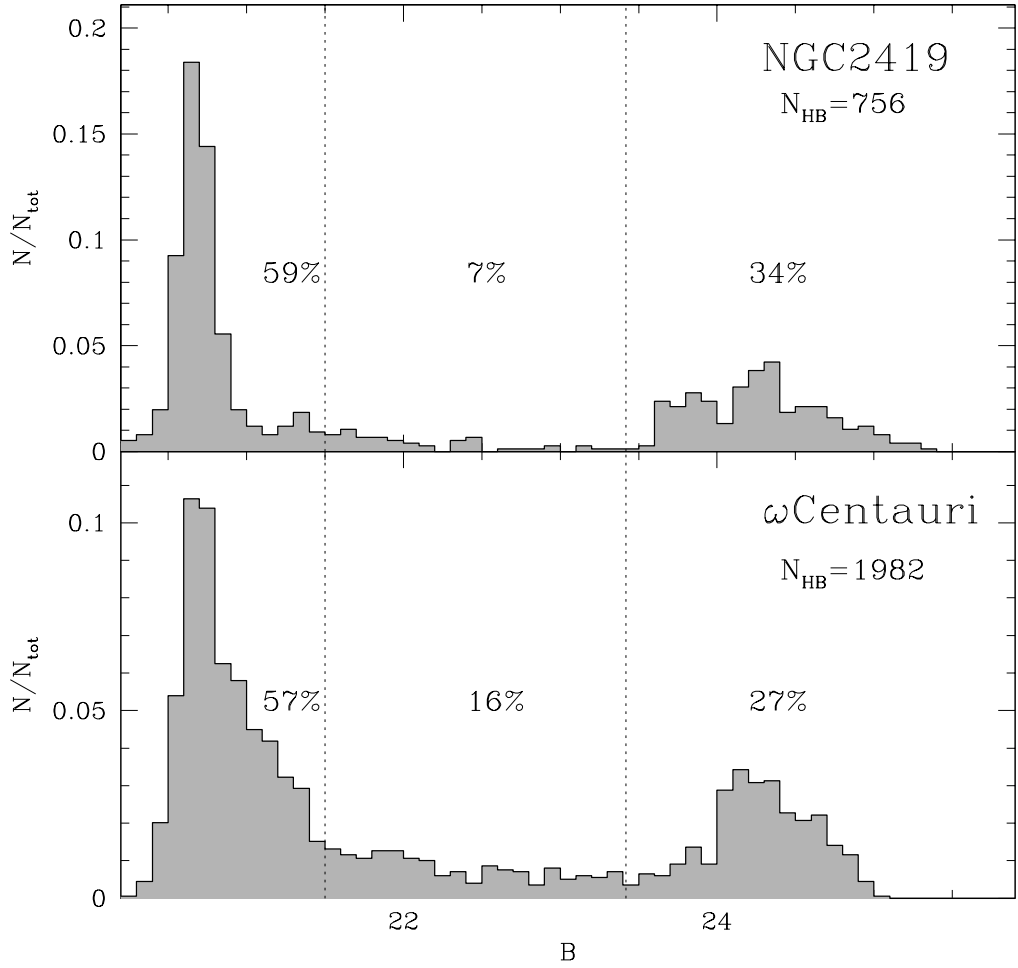}
\caption{Normalized magnitude distributions of the HB stars of
  NGC\,2419 (upper panel) and $\omega$ Cen (bottom panel, from F06)
  plotted in Fig.\,4. The vertical dotted lines mark three
  (arbitrary) portions of the HB separting the bulk of the population,
  the BT HB and the BHk.}
\label{hbdistr}
\end{center}
\end{figure}

As in previous papers (see e.g. Ferraro et al. 1995,
F06) here we compute the PB-BSS frequency as the number of BSS
normalized to the sampled luminosity in units of $10^4\,L_\odot$:
$S4_{\rm PB-BSS}= \nbss/L_4$ (see the bottom panel of
Fig.\,\ref{Npop}).  This quantity is useful for estimating the expected
number of BSS generated by PBs for each fraction of the sampled light
in any stellar system, resolved or not\footnote{This
    quantity is also important in evaluating the incidence of
    creation/destruction rate of BSS in the central region of
    high-density clusters, where collisions can have played a major
    role in producing collisional BSS.}.
For NGC\,2419, we find $S4_{\rm PB-BSS} =3.1 \pm 0.6$ (see
Fig.\,\ref{Npop}). Before comparing this quantity to that found in
$\omega$ Cen by F06, we must account for the fact that the adopted BSS
selection criteria are different in the two clusters.  In NGC\,2419 we
considered BSS brighter than $B<23.6$; this threshold corresponds to
$B<18$ at the distance of $\omega$ Cen, using the value $\delta B=5.6$
needed to shift the HB of $\omega$ Cen onto that of NGC\,2419 (see
Fig.\ref{hbdistr}).  By adopting this threshold, 104 BSS are found in
the ACS FoV of $\omega$ Cen, and by considering the sampled
luminosity, we obtain $S4_{\rm PB-BSS} = 1.6$ for this
cluster.\footnote{Of course, such a value is slightly smaller than
  that ($S4_{\rm PB-BSS} =2$) obtained in F06 by considering the
  entire sample of BSS with $B<18.4$} This comparison suggests that
the number of BSS per unit sampled light in NGC\,2419 is twice as
large than that in $\omega$ Cen.\footnote{A similar result is obtained
  if selecting the BSS population of NGC\,2419 down to the same
  threshold used by F06 for the BSS in $\omega$ Cen, i.e. $B<18.4$
  (which corresponds to $B=24$ at the distance of NGC\,2419).}  Under
the hypothesis that the vast majority of these BSS are generated by
the evolution of PBs, the different $S4_{\rm PB-BSS}$ values could
result from a different binary frequency in the two clusters, since
PB-BSS are expected to strongly depend on the fraction of binaries in
the cluster. Indeed the first direct correlation between these two
quantities (the binary fraction and the BSS frequency) has been
recently detected in a sample of 13 low-density clusters by Sollima et
al. (2008). Such a connection strongly supports a scenario in which
the evolution of PBs is the main formation channel for BSS in
low-density environments.
 
The case of NGC\,2419 further supports the idea that important
signatures of the dynamical evolution of the parent cluster are
imprinted in the radial distribution of the BSS population: indeed the
most recent results collected by our group (see Ferraro et al. 2003,
2006; Lanzoni et al. 2007a,b,c; Dalessandro et al. 2007) are building
the ideal data-base from which such signatures can be read and
interpreted, and are already confirming this hyphothesis.

\acknowledgements 
This research was supported by PRIN-INAF2006, by the Agenzia Spaziale
Italiana (under contract ASI I/088/06/0)
 and by the Ministero dell'Istruzione, dell'Universit\`a e
della Ricerca. It is also part of the {\it Progetti Strategici di
  Ateneo 2006} granted by the University of Bologna. ED is supported
by ASI; he acknowledges the {\it Marco Polo} programme of the
Bologna University for a grant support, and the Department of Astronomy
of the University of Virginia for the hospitality during his stay when
part of this work was done. RTR is partially supported by STScI grant
GO-10524.  This research has made use of the ESO/ST-ECF Science
Archive facility which is a joint collaboration of the European
Southern Observatory and the Space Telescope - European Coordinating
Facility.

\clearpage
\begin{deluxetable}{lccccc}
\tabletypesize{\footnotesize}
\tablewidth{0cm}
\tablecaption{The BSS population of NGC\,2419}
\tablehead{
\colhead{Name} &
\colhead{RA} &
\colhead{DEC} &
\colhead{B} &
\colhead{V} &
\colhead{I}\\
\colhead{ } & 
\colhead{[degree]} &
\colhead{[degree]} &
\colhead{ } &
\colhead{ } &
\colhead{ }
}
\startdata
BSS   1 & 114.5346238 &  38.8659223 & 21.67 & 21.48 & 21.19 \\
BSS   2 & 114.5317934 &  38.8778178 & 21.97 & 21.84 & 21.60 \\
BSS   3 & 114.5545324 &  38.8522824 & 22.01 & 21.86 & 21.55 \\
BSS   4 & 114.5290061 &  38.8781085 & 21.96 & 21.87 & 21.78 \\
BSS   5 & 114.5563785 &  38.8777071 & 22.03 & 21.88 & 21.67 \\
BSS   6 & 114.5375980 &  38.8849180 & 21.99 & 21.88 & 21.70 \\
BSS   7 & 114.5426076 &  38.8723283 & 22.20 & 21.98 & 21.72 \\
BSS   8 & 114.5322465 &  38.8824554 & 22.08 & 21.98 & 21.86 \\
BSS   9 & 114.5321660 &  38.8802393 & 22.10 & 21.99 & 21.81 \\
BSS  10 & 114.5314467 &  38.8844550 & 22.11 & 21.99 & 21.72 \\
BSS  11 & 114.5389799 &  38.8741541 & 22.20 & 22.04 & 21.83 \\
BSS  12 & 114.5180869 &  38.8707066 & 22.20 & 22.04 & 21.88 \\
BSS  13 & 114.5333262 &  38.8735631 & 22.24 & 22.09 & 21.62 \\
BSS  14 & 114.5186506 &  38.8555156 & 22.15 & 22.09 & 22.05 \\
BSS  15 & 114.5631273 &  38.8567423 & 22.34 & 22.10 & 21.72 \\
BSS  16 & 114.5350018 &  38.8575038 & 22.36 & 22.14 & 21.79 \\
BSS  17 & 114.5220854 &  38.8869184 & 22.32 & 22.14 & 21.83 \\
BSS  18 & 114.5559528 &  38.8756214 & 22.36 & 22.14 & 21.82 \\
BSS  19 & 114.5308496 &  38.8836661 & 22.36 & 22.15 & 21.86 \\
BSS  20 & 114.5052076 &  38.8695694 & 22.27 & 22.18 & 22.06 \\
BSS  21 & 114.5479750 &  38.8829417 & 22.39 & 22.23 & 21.98 \\
BSS  22 & 114.5542824 &  38.8787225 & 22.35 & 22.23 & 22.04 \\
BSS  23 & 114.5279917 &  38.8771869 & 22.33 & 22.23 & 22.05 \\
BSS  24 & 114.5289397 &  38.8789566 & 22.35 & 22.25 & 22.22 \\
BSS  25 & 114.5402618 &  38.8322954 & 22.45 & 22.30 & 21.80 \\
BSS  26 & 114.5544769 &  38.8807888 & 22.51 & 22.30 & 21.96 \\
BSS  27 & 114.5179647 &  38.8667478 & 22.53 & 22.35 & 22.06 \\
BSS  28 & 114.5208423 &  38.8864626 & 22.44 & 22.37 & 22.17 \\
BSS  29 & 114.5099740 &  38.8743802 & 22.49 & 22.38 & 22.18 \\
BSS  30 & 114.5707475 &  38.8723047 & 22.56 & 22.39 & 22.14 \\
BSS  31 & 114.5349532 &  38.8807603 & 22.62 & 22.43 & 22.14 \\
BSS  32 & 114.5373408 &  38.8847449 & 22.68 & 22.43 & 22.06 \\
BSS  33 & 114.5255999 &  38.8898043 & 22.69 & 22.46 & 22.23 \\
BSS  34 & 114.5213599 &  38.8819734 & 22.63 & 22.49 & 22.32 \\
BSS  35 & 114.5624518 &  38.8577883 & 22.68 & 22.51 & 22.22 \\
BSS  36 & 114.5502354 &  38.8587983 & 22.65 & 22.52 & 22.29 \\
BSS  37 & 114.5724785 &  38.8543265 & 22.70 & 22.52 & 22.22 \\
BSS  38 & 114.5377048 &  38.8776849 & 22.74 & 22.56 & 22.38 \\
BSS  39 & 114.5367933 &  38.8863131 & 22.75 & 22.56 & 22.25 \\
BSS  40 & 114.5370632 &  38.8841385 & 22.65 & 22.57 & 22.41 \\
BSS  41 & 114.5416913 &  38.8744374 & 22.80 & 22.57 & 22.20 \\
BSS  42 & 114.5328193 &  38.8783052 & 22.61 & 22.57 & 22.29 \\
BSS  43 & 114.5410755 &  38.8636039 & 22.70 & 22.58 & 22.35 \\
BSS  44 & 114.5443444 &  38.8841840 & 22.78 & 22.58 & 22.18 \\
BSS  45 & 114.5355871 &  38.8586443 & 22.72 & 22.59 & 22.29 \\
BSS  46 & 114.5330413 &  38.8827233 & 22.75 & 22.59 & 22.34 \\
BSS  47 & 114.5100504 &  38.8739744 & 22.78 & 22.59 & 22.38 \\
BSS  48 & 114.5394051 &  38.8758936 & 22.74 & 22.60 & 22.44 \\
BSS  49 & 114.5271347 &  38.8795152 & 22.76 & 22.61 & 22.42 \\
BSS  50 & 114.5277051 &  38.8748074 & 22.77 & 22.64 & 22.42 \\
BSS  51 & 114.5364186 &  38.8728954 & 22.88 & 22.65 & 22.22 \\
BSS  52 & 114.5358544 &  38.8787353 & 22.86 & 22.66 & 22.44 \\
BSS  53 & 114.4982285 &  38.8531324 & 22.78 & 22.67 & 22.14 \\
BSS  54 & 114.5145126 &  38.8789390 & 22.88 & 22.69 & 22.41 \\
BSS  55 & 114.5136075 &  38.8758341 & 22.84 & 22.70 & 22.50 \\
BSS  56 & 114.5436548 &  38.8769220 & 22.94 & 22.70 & 22.35 \\
BSS  57 & 114.5435517 &  38.8814084 & 22.92 & 22.70 & 22.40 \\
BSS  58 & 114.5381816 &  38.8832314 & 22.99 & 22.71 & 22.25 \\
BSS  59 & 114.5366860 &  38.8800032 & 22.88 & 22.72 & 22.39 \\
BSS  60 & 114.5285765 &  38.8761503 & 22.82 & 22.72 & 22.40 \\
BSS  61 & 114.5653514 &  38.8672278 & 22.90 & 22.74 & 22.41 \\
BSS  62 & 114.5319576 &  38.8456597 & 22.89 & 22.74 & 22.52 \\
BSS  63 & 114.5332823 &  38.8808548 & 22.90 & 22.75 & 22.50 \\
BSS  64 & 114.5201658 &  38.8716290 & 23.05 & 22.75 & 22.49 \\
BSS  65 & 114.5182917 &  38.8645682 & 22.93 & 22.76 & 22.43 \\
BSS  66 & 114.5422667 &  38.8810701 & 23.00 & 22.77 & 22.48 \\
BSS  67 & 114.5383605 &  38.8855947 & 23.02 & 22.77 & 22.42 \\
BSS  68 & 114.5228769 &  38.8715036 & 22.95 & 22.78 & 22.58 \\
BSS  69 & 114.5655782 &  38.8738971 & 22.95 & 22.79 & 22.54 \\
BSS  70 & 114.5508635 &  38.8556174 & 22.91 & 22.80 & 22.44 \\
BSS  71 & 114.5248251 &  38.8798819 & 23.03 & 22.80 & 22.50 \\
BSS  72 & 114.5330946 &  38.8782802 & 22.94 & 22.81 & 22.42 \\
BSS  73 & 114.5300473 &  38.8726778 & 22.99 & 22.82 & 22.42 \\
BSS  74 & 114.5258531 &  38.8844487 & 22.98 & 22.82 & 22.57 \\
BSS  75 & 114.5348296 &  38.8738382 & 23.04 & 22.84 & 22.40 \\
BSS  76 & 114.5148586 &  38.8769283 & 23.00 & 22.85 & 22.74 \\
BSS  77 & 114.5516168 &  38.8734430 & 23.02 & 22.85 & 22.62 \\
BSS  78 & 114.5466460 &  38.8781617 & 23.02 & 22.86 & 22.60 \\
BSS  79 & 114.5394030 &  38.8844577 & 23.08 & 22.86 & 22.43 \\
BSS  80 & 114.5282537 &  38.8767212 & 23.03 & 22.88 & 22.50 \\
BSS  81 & 114.5306900 &  38.8770293 & 23.08 & 22.89 & 22.55 \\
BSS  82 & 114.5385251 &  38.8771382 & 23.00 & 22.90 & 22.66 \\
BSS  83 & 114.5235814 &  38.8729527 & 23.16 & 22.92 & 22.48 \\
BSS  84 & 114.5355456 &  38.8739955 & 23.18 & 22.92 & 22.46 \\
BSS  85 & 114.5333547 &  38.8705373 & 23.05 & 22.93 & 22.76 \\
BSS  86 & 114.5268290 &  38.8888513 & 23.08 & 22.93 & 22.65 \\
BSS  87 & 114.5286345 &  38.8824567 & 23.17 & 22.95 & 22.57 \\
BSS  88 & 114.5272080 &  38.8829733 & 23.15 & 22.97 & 22.68 \\
BSS  89 & 114.5297549 &  38.8842508 & 23.17 & 22.97 & 22.74 \\
BSS  90 & 114.5374061 &  38.8851489 & 23.23 & 22.98 & 22.72 \\
BSS  91 & 114.5305231 &  38.8777378 & 23.19 & 22.99 & 22.63 \\
BSS  92 & 114.5306248 &  38.8655249 & 23.12 & 23.00 & 22.44 \\
BSS  93 & 114.5298495 &  38.8808827 & 23.18 & 23.00 & 22.55 \\
BSS  94 & 114.5250029 &  38.8861768 & 23.26 & 23.00 & 22.62 \\
BSS  95 & 114.5319072 &  38.8857620 & 23.20 & 23.01 & 22.75 \\
BSS  96 & 114.5766634 &  38.8573333 & 23.27 & 23.02 & 22.64 \\
BSS  97 & 114.5540530 &  38.8462298 & 23.17 & 23.03 & 22.62 \\
BSS  98 & 114.5446661 &  38.8786382 & 23.24 & 23.04 & 22.85 \\
BSS  99 & 114.5416242 &  38.8825834 & 23.19 & 23.04 & 22.57 \\
BSS 100 & 114.5559916 &  38.8782646 & 23.32 & 23.06 & 22.67 \\
BSS 101 & 114.5467694 &  38.8799242 & 23.34 & 23.07 & 22.71 \\
BSS 102 & 114.5245139 &  38.8773725 & 23.24 & 23.07 & 22.71 \\
BSS 103 & 114.5335661 &  38.8876156 & 23.24 & 23.07 & 22.85 \\
BSS 104 & 114.5338201 &  38.8808881 & 23.32 & 23.07 & 22.79 \\
BSS 105 & 114.5251903 &  38.8708786 & 23.30 & 23.07 & 22.63 \\
BSS 106 & 114.5155847 &  38.8515763 & 23.25 & 23.08 & 22.62 \\
BSS 107 & 114.5482144 &  38.8762432 & 23.24 & 23.09 & 22.69 \\
BSS 108 & 114.5283801 &  38.8804424 & 23.32 & 23.09 & 22.73 \\
BSS 109 & 114.5272202 &  38.8806563 & 23.33 & 23.09 & 22.75 \\
BSS 110 & 114.5329037 &  38.8787814 & 23.21 & 23.10 & 22.82 \\
BSS 111 & 114.5648250 &  38.8809186 & 23.39 & 23.10 & 22.67 \\
BSS 112 & 114.5655476 &  38.8353859 & 23.08 & 23.11 & 22.62 \\
BSS 113 & 114.5340609 &  38.8672425 & 23.38 & 23.11 & 22.72 \\
BSS 114 & 114.5241875 &  38.8826388 & 23.39 & 23.11 & 22.76 \\
BSS 115 & 114.5440484 &  38.8728605 & 23.33 & 23.12 & 22.70 \\
BSS 116 & 114.5318474 &  38.8804939 & 23.31 & 23.12 & 22.82 \\
BSS 117 & 114.4975565 &  38.8498301 & 23.36 & 23.14 & 22.69 \\
BSS 118 & 114.5308869 &  38.8858920 & 23.31 & 23.14 & 22.65 \\
BSS 119 & 114.5319236 &  38.8814812 & 23.39 & 23.14 & 22.77 \\
BSS 120 & 114.5365736 &  38.8749561 & 23.34 & 23.14 & 22.74 \\
BSS 121 & 114.5548031 &  38.8770840 & 23.36 & 23.15 & 22.84 \\
BSS 122 & 114.5698279 &  38.8632929 & 23.43 & 23.15 & 22.88 \\
BSS 123 & 114.5429103 &  38.8833364 & 23.26 & 23.15 & 22.81 \\
BSS 124 & 114.5180654 &  38.8526066 & 23.35 & 23.15 & 22.81 \\
BSS 125 & 114.5713938 &  38.8644595 & 23.41 & 23.15 & 22.81 \\
BSS 126 & 114.5151887 &  38.8881692 & 23.29 & 23.16 & 22.85 \\
BSS 127 & 114.4971710 &  38.8503894 & 23.39 & 23.18 & 22.80 \\
BSS 128 & 114.5389324 &  38.8662627 & 23.35 & 23.18 & 22.72 \\
BSS 129 & 114.5653667 &  38.8735289 & 23.35 & 23.19 & 22.80 \\
BSS 130 & 114.5290415 &  38.8850904 & 23.46 & 23.19 & 22.82 \\
BSS 131 & 114.5263335 &  38.8860022 & 23.29 & 23.20 & 22.80 \\
BSS 132 & 114.5704264 &  38.8781784 & 23.35 & 23.20 & 22.91 \\
BSS 133 & 114.5450942 &  38.8616024 & 23.42 & 23.21 & 22.90 \\
BSS 134 & 114.5261968 &  38.8738325 & 23.43 & 23.21 & 22.75 \\
BSS 135 & 114.5411974 &  38.8462718 & 23.46 & 23.21 & 22.85 \\
BSS 136 & 114.5459761 &  38.8828031 & 23.37 & 23.21 & 22.86 \\
BSS 137 & 114.5219796 &  38.8885754 & 23.44 & 23.25 & 22.97 \\
BSS 138 & 114.5550069 &  38.8277938 & 23.47 & 23.26 & 22.97 \\
BSS 139 & 114.5343111 &  38.8853578 & 23.49 & 23.27 & 22.86 \\
BSS 140 & 114.5707653 &  38.8713824 & 23.52 & 23.27 & 22.85 \\
BSS 141 & 114.5633208 &  38.8755186 & 23.52 & 23.28 & 22.91 \\
BSS 142 & 114.5372475 &  38.8697969 & 23.58 & 23.29 & 22.83 \\
BSS 143 & 114.5128112 &  38.8761214 & 23.54 & 23.30 & 22.95 \\
BSS 144 & 114.5303288 &  38.8832148 & 23.55 & 23.31 & 22.99 \\
BSS 145 & 114.5293403 &  38.8792390 & 23.56 & 23.31 & 22.81 \\
BSS 146 & 114.5403815 &  38.8621857 & 23.46 & 23.32 & 22.93 \\
BSS 147 & 114.5328409 &  38.8759082 & 23.58 & 23.34 & 23.02 \\
BSS 148 & 114.5058275 &  38.8390039 & 23.55 & 23.35 & 22.96 \\
BSS 149 & 114.5354601 &  38.8773111 & 23.55 & 23.38 & 22.94 \\
BSS 150 & 114.5289120 &  38.8672031 & 23.58 & 23.38 & 22.88 \\
BSS 151 & 114.5165945 &  38.8734688 & 23.51 & 23.39 & 23.01 \\
BSS 152 & 114.5364602 &  38.8814663 & 22.50 & 22.51 & 22.27 \\
BSS 153 & 114.5340522 &  38.8779415 & 22.62 & 22.62 & 22.42 \\
BSS 154 & 114.5347666 &  38.8829015 & 22.75 & 22.79 & 22.62 \\
BSS 155 & 114.5593850 &  38.8513222 & 22.80 & 22.76 & 22.40 \\
BSS 156 & 114.5358572 &  38.8710266 & 22.88 & 22.91 & 22.47 \\
BSS 157 & 114.5439332 &  38.8315146 & 22.97 & 22.98 & 22.51 \\
BSS 158 & 114.5550096 &  38.8666358 & 22.95 & 22.60 & 22.59 \\
BSS 159 & 114.5358955 &  38.8824612 & 23.24 & 22.88 & 22.63 \\
BSS 160 & 114.5736822 &  38.8698082 & 23.27 & 22.95 & 22.64 \\
BSS 161 & 114.5383375 &  38.8831536 & 23.28 & 22.89 & 22.53 \\
BSS 162 & 114.5363135 &  38.8768683 & 23.34 & 22.97 & 22.70 \\
BSS 163 & 114.5411630 &  38.8818491 & 23.37 & 22.99 & 22.73 \\
BSS 164 & 114.5385990 &  38.8814390 & 23.46 & 23.16 & 22.74 \\
BSS 165 & 114.5303744 &  38.8839579 & 23.53 & 23.22 & 22.90 \\
BSS 166 & 114.5342489 &  38.8869653 & 23.55 & 23.52 & 23.12 \\
BSS 167 & 114.5326253 &  38.8762184 & 23.56 & 23.18 & 22.82 \\
BSS 168 & 114.5274273 &  38.8465093 & 22.44 & 22.52 & 22.09 \\
BSS 169 & 114.5294680 &  38.8762898 & 21.67 &   -   & 21.58 \\
BSS 170 & 114.5590138 &  38.8529876 & 22.61 &   -   & 22.13 \\
BSS 171 & 114.5352678 &  38.8831155 & 23.27 &   -   & 22.54 \\
BSS 172 & 114.5293589 &  38.8824958 & 23.28 &   -   & 22.54 \\
BSS 173 & 114.5331303 &  38.8781672 & 22.37 &   -   & 21.82 \\
BSS 174 & 114.5372462 &  38.8804816 & 22.38 &   -   & 21.92 \\
BSS 175 & 114.5371350 &  38.8825563 & 23.11 &   -   & 22.68 \\
BSS 176 & 114.5299404 &  38.8833588 & 23.03 &   -   & 22.66 \\
BSS 177 & 114.5377141 &  38.8839299 & 22.28 &   -   & 21.96 \\
BSS 178 & 114.5318896 &  38.8816916 & 23.40 &   -   & 22.72 \\
BSS 179 & 114.5338975 &  38.8878113 & 22.30 &   -   & 21.76 \\
BSS 180 & 114.5354840 &  38.8848507 & 21.99 &   -   & 21.99 \\
BSS 181 & 114.5416767 &  38.8787643 & 22.16 &   -   & 21.86 \\
BSS 182 & 114.5274273 &  38.8465093 & 22.44 &   -   & 22.09 \\
BSS 183 & 114.5430587 &  38.8788861 & 22.52 &   -   & 22.27 \\
BSS 184 & 114.6046393 &  38.8739340 &   -   & 21.61 & 21.50 \\
BSS 185 & 114.4732159 &  38.8794161 &   -   & 22.22 & 22.11 \\
BSS 186 & 114.5948464 &  38.8317186 &   -   & 22.29 & 22.18 \\
BSS 187 & 114.4921101 &  38.8826003 &   -   & 22.36 & 22.13 \\
BSS 188 & 114.4795474 &  38.8950070 &   -   & 22.42 & 22.00 \\
BSS 189 & 114.5742838 &  38.8787606 &   -   & 22.51 & 22.11 \\
BSS 190 & 114.5769900 &  38.8517555 &   -   & 22.53 & 22.25 \\
BSS 191 & 114.5235357 &  38.9041986 &   -   & 22.57 & 22.32 \\
BSS 192 & 114.6028209 &  38.8986391 &   -   & 22.59 & 22.26 \\
BSS 193 & 114.5780235 &  38.8817962 &   -   & 22.63 & 22.39 \\
BSS 194 & 114.5205417 &  38.9003089 &   -   & 22.65 & 22.44 \\
BSS 195 & 114.5574956 &  38.8950661 &   -   & 22.68 & 22.37 \\
BSS 196 & 114.6020290 &  38.8716163 &   -   & 22.69 & 22.43 \\
BSS 197 & 114.4870399 &  38.8486762 &   -   & 22.73 & 22.39 \\
BSS 198 & 114.5278361 &  38.9056871 &   -   & 22.74 & 22.37 \\
BSS 199 & 114.4868008 &  38.8777003 &   -   & 22.75 & 22.48 \\
BSS 200 & 114.5698058 &  38.8856298 &   -   & 22.80 & 22.43 \\
BSS 201 & 114.5631611 &  38.9001108 &   -   & 22.80 & 22.45 \\
BSS 202 & 114.5728569 &  38.9152703 &   -   & 22.80 & 22.53 \\
BSS 203 & 114.5344124 &  38.9094217 &   -   & 22.82 & 22.45 \\
BSS 204 & 114.4987279 &  38.9233287 &   -   & 22.84 & 22.50 \\
BSS 205 & 114.4950878 &  38.9045856 &   -   & 22.90 & 22.61 \\
BSS 206 & 114.5542729 &  38.8926652 &   -   & 22.93 & 22.61 \\
BSS 207 & 114.4846796 &  38.8934777 &   -   & 22.96 & 22.63 \\
BSS 208 & 114.4793021 &  38.8865969 &   -   & 22.96 & 22.81 \\
BSS 209 & 114.4748604 &  38.9055783 &   -   & 22.98 & 22.61 \\
BSS 210 & 114.5080023 &  38.9026343 &   -   & 22.97 & 22.72 \\
BSS 211 & 114.5573887 &  38.8903576 &   -   & 22.98 & 22.63 \\
BSS 212 & 114.5204611 &  38.9028901 &   -   & 23.08 & 22.85 \\
BSS 213 & 114.5705620 &  38.8861123 &   -   & 23.09 & 22.81 \\
BSS 214 & 114.5548202 &  38.9085582 &   -   & 23.11 & 22.72 \\
BSS 215 & 114.5422886 &  38.8977372 &   -   & 23.12 & 22.84 \\
BSS 216 & 114.4864816 &  38.8823681 &   -   & 23.16 & 22.77 \\
BSS 217 & 114.5658903 &  38.8861262 &   -   & 23.17 & 22.84 \\
BSS 218 & 114.5066897 &  38.8939685 &   -   & 23.20 & 22.79 \\
BSS 219 & 114.4888325 &  38.8881818 &   -   & 23.27 & 22.80 \\
BSS 220 & 114.5077125 &  38.9123421 &   -   & 23.27 & 22.81 \\
BSS 221 & 114.4809893 &  38.8830761 &   -   & 23.27 & 23.04 \\
BSS 222 & 114.5560389 &  38.8862473 &   -   & 23.29 & 23.07 \\
BSS 223 & 114.4463878 &  38.8771839 &   -   & 22.20 & 22.10 \\
BSS 224 & 114.3957223 &  38.8532619 &   -   & 22.65 & 22.27 \\
BSS 225 & 114.4183885 &  38.8968389 &   -   & 22.86 & 22.52 \\
BSS 226 & 114.4386092 &  38.8140150 &   -   & 22.90 & 22.59 \\
BSS 227 & 114.3704635 &  38.8294846 &   -   & 23.09 & 22.73 \\
BSS 228 & 114.5111720 &  38.9412987 &   -   & 22.61 & 22.43 \\
BSS 229 & 114.5357260 &  39.0140900 &   -   & 22.74 & 22.47 \\
BSS 230 & 114.4848049 &  38.9588305 &   -   & 22.99 & 22.67 \\
BSS 231 & 114.7046122 &  38.8578500 &   -   & 22.18 & 21.92 \\
BSS 232 & 114.6843307 &  38.9221184 &   -   & 23.28 & 22.83
\enddata
\label{tab:bss}
\end{deluxetable}

\clearpage
\begin{deluxetable}{rrrrrrrc}
\footnotesize
\tablecaption{Number Counts of BSS, HB, and RGB Stars, and Fraction of
Sampled Luminosity}
\tablewidth{11cm}
\tablehead{
\colhead{$r_i\arcsec$} &
\colhead{$r_e\arcsec$} &
\colhead{ } &
\colhead{$N_{\rm BSS}$} &
\colhead{$N_{\rm HB}$} &
\colhead{ } &
\colhead{$N_{\rm RGB}$} &
\colhead{$L^{\rm samp}/L_{\rm tot}^{\rm samp}$}
}
\startdata
   0 &   20 && 56 ~~~   & 160 ~~~     &&  745~~~~~ & 0.21 \\
  20 &   60 && 71 ~~~   & 253 ~~~     && 1137~~~~~ & 0.31 \\
  60 &  100 && 41 ~~~   & 142 ~~~     &&  592~$\,$(1) & 0.21 \\
 100 &  180 && 43$\,$(1)& 121$\,$(1)  &&  497~$\,$(3) & 0.18 \\
 180 &  500 && 15$\,$(5)&  77$\,$(11) &&  248$\,$(27) & 0.09 \\
\hline 
\enddata
\tablecomments{The values listed out of the parenthesis correspond to the
number of stars assumed to belong to the cluster (and thus used in the analysis),
while those in the parenthesis are estimated to be contaminating field
stars (see Sect. \ref{sec:radBSS}).}
\label{tab:counts}
\end{deluxetable}

\clearpage

\end{document}